\documentclass[aps,prb,preprintnumbers,twocolumn,
 showpacs,secnumarabic]{revtex4}

\usepackage{graphicx,setspace}
\usepackage{amsfonts,amsmath,amssymb,bm}
\usepackage[dvips,usenames]{color}

\begin{document}

\title{Confinement effects on diffusiophoretic self-propellers}

\date{\today}

\author{M. N. Popescu}
\email{Mihail.Popescu@unisa.edu.au}
\altaffiliation{also at:
Max-Planck-Institut f\"ur Metallforschung, Heisenbergstr. 3,
70569 Stuttgart, Germany}
\affiliation{
Ian Wark Research Institute, University of South Australia,
5095 Adelaide, South Australia, Australia}
\author{S. Dietrich}
\email{dietrich@mf.mpg.de}
\affiliation{
Max-Planck-Institut f\"ur Metallforschung, Heisenbergstr. 3,
70569 Stuttgart, Germany,}
\affiliation{Institut f\"ur Theoretische und Angewandte Physik,
Universit\"at Stuttgart, Pfaffenwaldring 57, 70569 Stuttgart,
Germany}
\author{G. Oshanin}
\email{oshanin@lptl.jussieu.fr}
\altaffiliation{also at: Max-Planck-Institut f\"ur Metallforschung,
Heisenbergstr. 3, 70569 Stuttgart, Germany}
\affiliation{Laboratoire de Physique Th{\'e}orique de la Mati{\`e}re
Condens{\'e}e, Universit{\'e} Pierre et Marie Curie (Paris 6) -
4 Place Jussieu, 75252 Paris, France}

\begin{abstract}
We study theoretically the effects of spatial confinement on the
phoretic motion of a dissolved particle driven by composition
gradients generated by chemical reactions of its solvent, which
are active only on certain parts of the particle surface.
We show that the presence of confining walls increases in a
similar way both the composition gradients and the viscous
friction, and the overall result of these competing effects is an
increase in the phoretic velocity of the particle. For the case of
steric repulsion only between the particle and the product molecules
of the chemical reactions, the absolute value of the velocity remains
nonetheless rather small.
\end{abstract}


\maketitle

\section{Introduction}
\label{intro} Recent years have witnessed a growing technological,
experimental, and theoretical interest in scaling standard machinery
down to micro- and nano-scales as needed for the development of
``lab on a chip'' devices. For applications in, e.g., drug-delivery
systems or micromechanics one of the most challenging problems at
this stage is to develop ways to enable small-scale objects to
perform autonomous, controlled motion \cite{Paxton_2005,Paxton_2006}.
Although the research in this area is still in its early stages,
several such proposals have already been tested experimentally
(see, e.g.,
Refs. \onlinecite{Whitesides_2002,Sen_2005,Golestanian_2005,Howse_2007}).
A review of the recent progress in this field can be found in
Ref. \onlinecite{Paxton_2006}.

Whitesides and co-workers proposed a design of self-propelling
devices based on an asymmetric decoration of the surface of small
objects by catalytic, active sites promoting a chemical reaction in
the surrounding liquid medium \cite{Whitesides_2002}.
This asymmetric distribution can provide motility through a variety
of mechanisms, such as surface tension gradients and/or cyclic
adsorption and desorption \cite{Paxton_2006,Kapral_2007}.
The use of an asymmetric surface distribution of a catalyst has been
further proposed for an autonomous diffusiophoretic motion emerging
as a result of self-created density gradients
\cite{Golestanian_2005}. An experimental realization of such
systems, using platinum coated polystyrene spheres, has been
recently reported \cite{Howse_2007}. The issue of designing optimal
distributions of a catalyst for spherical and rod-like particles
has also been approached \cite{Golestanian_2007}.

For most of the applications in biological systems or in 'lab on a
chip'-type devices one has to deal with a complicated internal
structure of the systems, such as networks of narrow channels or
pores and various impenetrable impurities. In some situations, one
may even encounter a quasi two- or one-dimensional behavior, like
in the cases of bacteria motion on planar nutrient
substrates \cite{Peruani_2007} and of the motion of small particles
within a biological membrane \cite{Saffman_1975} or in a monolayer
adsorbed on a three-dimensional (\textit{3d}) liquid subphase
\cite{Joanny_1999}. Thus spatial confinement is a relevant feature
so that the assumption of the presence of an unconfined \textit{3d}
bulk reactive solvent \cite{Paxton_2005,Paxton_2006,Golestanian_2005}
may break down. Intuitively, one expects that spatial confinement
does influence the resulting motion of self-propelling objects like
the ones discussed above. But {\it a priori} it is not clear if
such effects are significant.

Starting from the model used in Ref. \onlinecite{Golestanian_2005}, here
we study the effects of spatial confinement on the phoretic motion
of a particle that generates number density gradients of the
product molecules emerging from the chemical reactions in the
depleting Newtonian liquid solvent. We shall focus on the simple
case in which the particle and the \textit{3d} solution of solvent
and product molecules are bounded by a spherical shell because (i)
this simple geometry allows for an exact solution and (ii) it has
been experimentally shown for a variety of phoretic systems that
confinement effects are dominated by the smallest confining length
scale \cite{Anderson_1989}. Thus this geometry has a paradigmatic
character.

Similarly to the earlier studies
\cite{Paxton_2005,Paxton_2006,Sen_2005,Golestanian_2005,Howse_2007},
our work is based on adopting the standard theory of phoresis for
the present case, in which the gradients are self-generated rather
than being produced and maintained by external sources. In doing so,
one is bound to make a number of assumptions that are either already
present in the classical theory of phoresis, or arise as a result
of mapping the description of such ``active'' surface particles onto
the framework of a theory developed to describe the case of inert
particles immersed in a pre-defined, externally controlled
concentration gradient. Since in the literature these assumptions are
often overlooked or not spelled out explicitly, we consider it as
necessary to provide also a critical discussion of the significant
assumptions involved by this mapping, as well as of some of those
assumptions implicitly contained in the standard theory of phoresis.
Accordingly, the outline of this paper is as follows. In Section
\ref{Model} we define the model and discuss some general aspects of
systems with self-generated motion, with particular emphasis on the
assumptions involved in adopting the standard theory of phoresis.
Section \ref{sec_diff_phor_velocity} is devoted to the computation
of the diffusiophoretic velocity. This includes the calculation of
the diffusiophoretic slip velocity and the determination of the
phoretic hydrodynamic flows and density profiles of the product
molecules around a self-propelling particle. The results are
discussed in Section \ref{discussion}, and we conclude in Section
\ref{summary} with a brief summary of our results and general
conclusions. Important details of our calculations are presented in
the Appendices \ref{app_A}, \ref{app_B}, and \ref{app_C}.

\section{The Model}
\label{Model}

\subsection{General aspects.}
\label{general}

\begin{figure}[!htb]
\includegraphics[width=.7\linewidth]{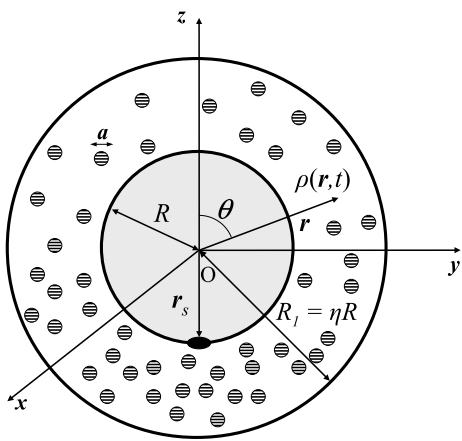}
\caption{
\label{fig1}
An impermeable, spherical particle of radius $R$ with a point-like
catalytic site at ${\mathbf{r}}_s = -R \mathbf{e}_z$ (depicted as
a black dot) on its surface, enclosed by a concentric, impermeable,
spherical wall of radius $R_1 = \eta R$ ($\eta > 1)$. The product
molecules of diameter $a$ are shown as small horizontally hatched
circles.}
\end{figure}

The system we consider is shown in Fig.~\ref{fig1}. It consists of
an impermeable, spherical particle of radius $R$ with a point-like
catalytic site (black dot in Fig.~\ref{fig1}) on its surface, which
promotes the chemical conversion of a surrounding solvent into
product molecules of diameter $a$ (small hatched circles in
Fig.~\ref{fig1}). The particle and the surrounding solution (solvent
plus the solute, i.e., the product molecules) of viscosity $\mu$ are
enclosed in a concentric, impermeable, spherical shell of radius
$R_1 = \eta R$ ($\eta > 1)$.

In general, such a chemical conversion of the solvent gives rise
to several types of product molecules. Here we shall focus on the
particular case in which the chemical conversion of a solvent
molecule ($A$) leads to two molecules only ($A'$ and $B$), one very
similar in size and properties with the solvent itself
($A' \approx A$), the other one ($B$) different
[$A \stackrel{cat}{\to} A' + B$]; in the following this latter is
denoted as ``product molecule'' and plays the role of a solute in the
solvent. In other words, we consider a situation in which the net
result of the chemical conversion can be approximated as the
generation of a solute, solely, and the reaction does not lead to a
solvent depletion near the catalytic site which otherwise would
reduce the current of solvent towards the catalytic site acting as
a solvent sink. For example, this is approximately the case for the
Pt catalyzed decomposition of hydrogen peroxide (H$_2$O$_2$) in
aqueous solution into water (H$_2$O) and oxygen (O$_2$) molecules,
as discussed in Refs. \onlinecite{Paxton_2005,Paxton_2006,Howse_2007}.
In these experimental studies the oxygen plays the role of the
product molecule the properties of which differ significantly from
those of the solvent. Here the solvent is actually a binary liquid
mixture of H$_2$O and H$_2$O$_2$ for which H$_2$O is passive and
does not participate in the chemical conversion. We
note that actually it is rather difficult to assess whether or
not H$_2$O and H$_2$O$_2$ can be treated as being the same in
such systems, and probably the answer would be on a case by case
basis. This is not only because the H$_2$O-particle and H$_2$O$_2$-particle interactions may be quite different for
different materials (the simplest example is exactly the catalytic decomposition of H$_2$O$_2$ on Pt), but also because of the
sensitivity of phoresis to the details of the \textit{solvent
mediated} interactions -- and the weakly acidic nature of
H$_2$O$_2$ may play an important role here.
In providing this example as a possible realization of a catalyzed reaction $A \stackrel{cat}{\to} A + B$ we had to rely on: (i) the statement
in Ref. \onlinecite{Paxton_2006} [second entry, bottom of left column
on page 13427 therein], which is based on some previously published results by Phibbs and Gigu{\`e}re \cite{Phibbs_1951}, that hydrogen peroxide and water in contact with Au have almost identical interfacial tensions (this translates into similar particle-solvent interactions), and thus that there is no additional solvent-density gradient to be considered,
and (ii) on the implicit statement in the same Ref. \onlinecite{Paxton_2006} [second entry, Eq. (8) therein] that the effective interaction of the
O$_2$ molecule with the Au surface of the rod does not depend on the composition of the solvent H$_2$O - H$_2$O$_2$ mixture.

We thus assume that the reaction at the catalytic site, (located at
${\mathbf{r}}_s = - R  \mathbf{e}_z$, where $\mathbf{e}_z$ is the
unit vector of the $z$-axis), acts effectively only as a point-like
source of product molecules \cite{Golestanian_2005} of diameter $a$
which are diffusing in the solvent with diffusion coefficient $D$.
In passing we note that this is in contrast to the situation
considered in Ref. \onlinecite{Kapral_2007}, where the catalytic site
acts both as a source for the product and as a sink for the solvent,
i.e., the generation of a product molecule is accompanied by the
annihilation of a solvent one so that $A \stackrel{cat}{\to} B$ which implies
that $B$ can only be a different configuration of $A$, because it
must have the same chemical constituents. In the situation
considered here, the amplitude of the production rate of $B$ is
denoted as $B(t)$. We shall neglect any interaction between the
product molecules. Thus the number density of product molecules is
considered to be so low that among themselves they behave like an
ideal gas. There is an interaction potential between the product
molecules and the moving particle (see Appendix \ref{app_A}),
which includes the impermeability condition at the surface of
particle. The interactions between the product molecules and the
solvent are accounted for in an effective way via the
Stokes - Einstein expression $D = k_B T/(3 \pi \mu a)$ for the
diffusion coefficient $D$ of the product molecules
\cite{S-E relation}, where $k_B$ is the Boltzmann constant and
$T$ is the temperature.

\subsection{Discussion of the model in the context of the standard
theory of phoresis.}
\label{sec_standard_phoresis}

The presence of a source of solute (product molecules) on the
surface of the particle creates a non-uniform and time dependent
distribution of solute in the solution (see Fig. \ref{fig1}),
i.e., a non-uniform composition of the solution. Because the
understanding of the way in which such a non-uniform composition
gives rise to phoretic motion is not straightforward, we discuss
here in some detail how the general model introduced in Section
\ref{general} can be put into the context of the standard theory
of phoresis \cite{Anderson_1989}. A recent clear exposition of
the basic concepts and a discussion of different phoretic transport
scenarios can be also found in Ref. \onlinecite{Prost}.

If the production rate of the source is not very large and the
diffusion coefficient of the product molecules through the solvent
is not too small, the solute density varies smoothly in space and
slowly in time. This justifies the assumption of local equilibrium
and the definition of a position- and time-dependent chemical
potential of the solution (the spatial gradients of which describe
the diffusion of the solute away from the source). In a quasi
steady-state of the solution, corresponding to slow time-variations
of the composition, the net solute current is approximately zero
and the solute density gradients are balanced by pressure gradients
[see also Eq. (B3) in Ref. \onlinecite{Prost}]. Because in the absence of
external body forces the mechanical equilibrium for a liquid
solution is generally established much faster than the chemical one
(see, e.g., Ref. \onlinecite{deGroot_book}), it is physically plausible
to assume that \textit{far from the boundaries} (i.e., far from the
wall and far from the particle surface) the pressure adjusts
instantaneously to accommodate the spatially varying solute density
profile \cite{Prost} and keeps varying slowly with time (on time
scales much larger than the diffusion time $\sim R^2/D$) because
the overall density of the solution increases in time. In other
words, the pressure is determined by the solute density profile
and is obtained from the equation of state of the solvent-solute
binary mixture considered as a thermodynamic system in local
equilibrium described by the corresponding free energy density.
Clearly, the situation is different in the region of contact between
the solution and the particle surface (and also of contact with the
wall) where the interactions of the particle with the product
molecules (solute) and with the solvent molecules are relevant. This
interaction disturbs the distribution of solute molecules, net
currents of solute and solvent result, and the pressure field
becomes a quantity determined by the incompressibility requirement
for the hydrodynamic flow, and not by an equation of state. We shall
discuss this important point below.

At a given time the spatial variations of the number density $\rho$
of the solute and of the number density $\rho_{solv}$ of the solvent
are characterized as follows: $\rho_{solv}$ is constant throughout
the solution apart from the close vicinity of the particle surface
and of the surface of the confining sphere. There $\rho_{solv}$
vanishes and reaches its constant bulk value via density oscillations
which are induced by local packing effects caused by the finite
diameter $a_{solv}$ of the solvent particles. Away from any kind of
surface phase transition the approach of the bulk value occurs
exponentially on the scale of the bulk correlation length
$\xi_{solv}$ of the solvent. (The presence of long-ranged dispersion
forces causes power-law decays; however, their amplitude is small
and is neglected in the present context.) Away from bulk phase
transitions  $\xi_{solv}$  is comparable with the range of the
interaction potential between the solvent molecules, which in turn
is proportional to $a_{solv}$ with a prefactor of order unity. Within
this picture $\rho_{solv}$ does not vary along the surface of the
particle.

The solute number density $\rho$ is characterized by two important
features. On the scale of the system size $R \gg a$, $\rho$ varies
due to the diffusion process. On the much smaller length scale of
the solute diameter $a$, $\rho$ also varies near the particle
surface and near the surface of the confinement as discussed above
for the solvent. Since, however, according to our earlier assumption
the solute particles can be considered to form an ideal gas, near
the walls $\rho$ varies proportional to $\exp( - \beta \Psi)$, where
$\Psi$ is the \textit{effective} substrate potential
(see Appendix \ref{app_A}) in the sense that it describes the
interaction of the solute molecule with the substrate in the presence
of the solvent, and $\beta = 1/k_B T$. Typically the range $\delta$
of $\Psi$ is proportional to the solute diameter $a$. Accordingly
the solute molecules interact directly with the particle only if
they are within a thin surface film of thickness $\delta$, which
is assumed to not be deformed by the motion of the particle
\cite{Anderson_1989}.

The standard theory of phoresis assumes $a,\,\delta \gg a_{solv}$ so
that within this surface film of thickness $\delta$ the solvent can
be considered as a continuum with constant $\rho_{solv}$. We note
that this assumption is rather convenient from a computational point
of view. However, this assumption is at odds with the basic
underlying model according to which the solute particle is created
from a solvent particle via a catalytic reaction. Under these
circumstances one expects $a \approx a_{solv}$ (see the above example
of the reaction $2$ H$_2$O$_2$ $\stackrel{cat}{\to}$ $2$ H$_2$O + O$_2$). In
the absence of a more detailed theory of phoresis, which treats the
sizes of the solvent and solute particles on the same footing, we
proceed along the lines of the standard theory. Nonetheless we point
out that there is an urgent need for improvement. It might be that
due to the fact that continuum hydrodynamics remains quantitatively
reliable down to surprisingly small length scales (here $a_{solv}$),
it is conceivable that the aforementioned continuum description
yields reliable results, too. But this remains unproven.

According to the standard theory a hydrodynamic description applies
within the aforementioned surface film. Within this picture the
solute molecules and their effective interaction with the particle
are replaced by a corresponding distribution of ``point forces''
acting on the solvent in the film. Within the limitations of such an
approach the hydrodynamic description of the solution naturally
splits into that of an ``inner'' region formed by the surface film
and that of the ``outer'' region formed by the exterior space beyond
the surface film. Following Refs. \onlinecite{Anderson_1989,Ajdari_2006,Golestanian_2007}, the ensuing
asymmetric, non-uniform solute number density $\rho(\mathbf{r},t)$
around the  particle will give rise, \textit{within the surface film
only}, to a pressure gradient along the surface of the particle. This
is because the surface film is very thin on the scale of the system
size $R$ so that the equilibration of the composition profile of the
solution within the surface film can be assumed to be fast along the
direction normal to the surface of the self-propelling particle
compared with the diffusional relaxation time of the composition
gradient along its surface, which typically involves a length scale
of the order of the particle size $R$. Therefore, within the surface
film the solute density in the direction normal to the surface of the
particle is given by a Boltzmann distribution corresponding to the
local equilibrium configuration in the presence of the effective
interaction potential between the particle and the solute molecules,
with a prefactor which depends on the position along the surface of
the particle (for details see Appendix \ref{app_A}, which builds on
Ref. \onlinecite{Anderson_1989}). Mechanical equilibrium of the solvent
within the surface film along the direction normal to the surface of
the particle (no flow along this direction) requires the pressure
gradient along the normal to be equal to the body force densities
due to the effective particle-solute interactions. Therefore the
pressure within the surface film differs from the ``outer'' pressure
field by an ``osmotic pressure'' term, i.e., a term proportional
to the extra solute density $\rho$ in excess to the constant solvent
density $\rho_{solv}$. Since this osmotic pressure varies along the
surface of the particle, there is a gradient of pressure along the
surface of the particle. This lateral pressure gradient is not
balanced by any body force and thus generates shear stress within
the surface film. Therefore it induces hydrodynamic flow of the
solution along the surface of the particle and entails motion of
the particle with a velocity $\mathbf{V}(t)$. Because the system
has azimuthal symmetry, the motion is along the $z$-axis, i.e.,
$\mathbf{V} = V \mathbf{e}_z$. The hydrodynamic flow of the
solution, the diffusive transport of the solute, and the coupling
between the two giving rise to phoretic motion of the particle are
analyzed in detail in the next section (see also the
Appendices \ref{app_A}-\ref{app_C}).

\section{Computation of the Diffusio-Phoretic Velocity}
\label{sec_diff_phor_velocity}

Our calculation of the diffusiophoretic velocity proceeds along
the lines of Ref. \onlinecite{Anderson_1989} and is based on dividing
the problem into an inner one -- within the film around the particle
surface discussed in Subsec. \ref{sec_standard_phoresis}, and an
outer one -- beyond the range of the effective interaction between
the solutes and the particle. In view of the arguments presented
in Sec. \ref{Model}, we base our analysis on the following
assumptions:\newline
(i) The chemical reaction leads to a change in the solute density
only. The thickness of the surface film, defined by the range of
the effective interaction between the product molecules (solute)
and the particle, is much smaller than the particle radius $R$.
The spatial variations of the solute number density along the
surface of the particle occur over length scales of the order of
$R$, which allows one to use the approximation of a locally
planar interface.
\newline
(ii) The number density of product molecules is sufficiently low,
so that the solute can be viewed as an ideal gas and the
solvent-solute mixture behaves like an ideal dilute solution.
In this case, the pressure gradient is simply proportional to the
gradient of the number density of the product particles (see
Appendix \ref{app_A}).
\newline
(iii) Temporal variations of the number density of the product
molecules due to their creation occur on time scales much longer
than those needed for the fluid flow (as seen from the moving
particle) to relax to a steady state corresponding to the number
density profile at that moment.
\newline
(iv) The flow field of the solution within the surface film can be
described by the laws of hydrodynamics.

Additionally, we assume that both the Reynolds number
$\mathrm{Re} \simeq \tilde \rho_{solv} V R/\mu$, where
$\tilde \rho_{solv}$ is the \textit{mass density} of the solvent,
and the Peclet number $\mathrm{Pe} \simeq V R/D$ are small, such
that one can approximate the hydrodynamic description with the
creeping flow (Stokes) equations and disregard the convection of
the solute compared to its diffusive transport.
Here we have assumed that the magnitude of the hydrodynamic flow
$\mathbf{u}$ is similar to that of the phoretic velocity
$V$; this assumption is supported \textit{a posteriori} by the fact
that for our system the phoretic velocity is basically the average
of the slip-velocity over the surface of the particle [see, c.f.,
Eqs. (\ref{slip_vel}) and (\ref{velocity})]. For a $\mu$m size
particle moving through water (density $\tilde \rho_{solv} =
10^3~\mathrm{kg/m}^3$, viscosity $\mu = 10^{-3}~\mathrm{Pa~s}$) with
a velocity of the order of $\mu$m/s, which is typical for phoresis,
one has $\mathrm{Re} \simeq 10^{-6}$. For the diffusion at room
temperature ($k_B T_{room} \sim 4 \times 10^{-21}$ J) of O$_2$
($a \sim 10^{-10}$ m) in H$_2$O$_2$ ($\mu \simeq
10^{-3}~\mathrm{Pa~s}$), the Stokes-Einstein relation leads to an
estimate $D \sim 4 \times 10^{-9} \mathrm{m}^2/\mathrm{s}$ for the
diffusion coefficient (in agreement with Ref. \onlinecite{Paxton_2006}),
and thus $\mathrm{Pe} \simeq 10^{-3}$. Therefore the latter
assumptions are justified. Note that if one uses $R_1$ rather than
$R$ as a characteristic length scale, the above results imply that
the $\mathrm{Re}$ and  $\mathrm{Pe}$ numbers remain both very small
as long as $\eta \lesssim 10$.

\subsection{Diffusiophoretic slip-velocity}
As discussed in Sec. \ref{Model}, the pressure gradient along the
surface of the particle, induced by its interaction with the
non-uniform distribution $\rho(\mathbf{r},t)$ of product molecules,
leads to flow of the solution relative to the particle. As shown in
Appendix \ref{app_A}, the hydrodynamic flow within the surface film
translates into a (phoretic) slip-velocity,
\begin{equation}
\label{slip_vel}
\mathbf{v}_s (\mathbf{r},t) = - b \nabla_s \rho(\mathbf{r},t)\,,
\textrm{ for } |\mathbf{r}| = R_+\,,
\end{equation}
at the outer edge $R_+ \gtrsim R$ of the surface film as a boundary
condition for the hydrodynamic flow in the outer region. In this
equation $\nabla_s$ denotes the projection of the gradient operator
onto the tangential planes of the surface of the particle,
\begin{equation}
\label{surf_grad}
\nabla_s =
\mathbf{e}_\theta \,\dfrac{1}{r}\dfrac{\partial}{\partial \theta}
+ \mathbf{e}_\phi \,
\dfrac{1}{r \sin \theta}\dfrac{\partial}{\partial \phi} \,,
\end{equation}
where $\mathbf{e}_\theta$ and $\mathbf{e}_\phi$ are the polar
and azimuthal unit vectors, respectively, while
\begin{equation}
\label{b_def}
b = \dfrac{k_B T}{\mu} \Lambda
\end{equation}
is an effective ``mobility'', and $\lambda = \sqrt{|\Lambda|}$ is
a characteristic length scale. The latter is determined by the
effective interaction potential $\Psi$ between the particle and the
product molecules \cite{Anderson_1989} [which determines their
distribution within the surface film along the direction $\hat y$
normal to the particle surface within the local coordinate system
(Fig. \ref{fig3}) or the radial direction here]:
\begin{equation}
\label{lambda_def}
\Lambda = \int\limits_0^\infty \, d \hat y\, \hat y
\left(e^{-\beta \Psi(\hat y)} - 1\right)\,.
\end{equation}
In the case of a purely steric repulsive interaction, one has \cite{Anderson_1989} $\Lambda = - a^2/8$ so that
$\lambda = a/(2\sqrt{2})$. For a single catalytically active site as
shown in Fig. \ref{fig1} $\rho$ and thus $\mathbf{v}_s$ do not
depend on the azimuthal angle $\phi$.

\subsection{Phoretic hydrodynamic flow}
\label{subsec_phor_flow}

We focus here on the quasi-static approximation (iii) described
above and the limit of low Re numbers. In the laboratory frame,
in which the particle moves with a yet unknown velocity
$\mathbf{V}(t;\eta)$, the hydrodynamic flow $\mathbf{u} \equiv
\mathbf{u}(\mathbf{r};t,\eta)$ at time $t$ in the domain beyond the
surface film around the particle obeys the steady-state force-free
Stokes equations
\begin{equation}
\label{St_eq}
\nabla \cdot \mathbf{\hat \Pi} = 0,
~\nabla \cdot \mathbf{u} = 0,
~~R_+ < |\mathbf{r}| < R_1\,,
\end{equation}
where $\mathbf{\hat \Pi} = -p \mathbf{\hat I} +
\mu \mathbf{\hat \Sigma}$ is the pressure tensor, $p$ is the
hydrostatic pressure, and $\mathbf{\hat \Sigma}$ is the shear
stress tensor, i.e., $\Sigma_{\alpha \beta} =
\partial u_{\alpha}/\partial x_{\beta} +
\partial u_{\beta}/\partial x_{\alpha}$, subject to boundary
conditions (BC) of no-slip at the wall at $R_1$ and prescribed slip
velocity on the surface  $R_+$, i.e.,
\begin{equation}
\label{BC_flow}
\left.\mathbf{u}\right|_{|\mathbf{r}|= R_+} =
\mathbf{V}(t;\eta) + \mathbf{v}_s,~~
\left.\mathbf{u}\right|_{|\mathbf{r}|= R_1} = 0\,.
\end{equation}
Note that the parametric dependences of the flow field
$\mathbf{u}$ on $t$ and $\eta$ stem from the boundary conditions.
Note that because for the outer problem the variations of the flow
field and those of the number density of the product molecules are
over length scales that are much larger than $\delta$, one can
replace everywhere in the calculations the (unknown) value of
$R_+$ by $R$; we shall use this approximation in the following.

The solution of the Stokes equations with boundary conditions on
spherical surfaces is based on expressing both the flow field
$\textbf{u}$ and the pressure field $p$ as series of solid harmonics \cite{Happel_book,Lamb_book} $K_l(r,\theta)$ ($l \in \mathbb{Z}$),
which are the eigenfunctions of the \textit{3d} Laplace operator. In
Appendix \ref{app_B} we provide a brief outline of the general
method for solving Eq. (\ref{St_eq}) in spherical coordinates and
derive the solution obeying the BCs given by Eq. \ref{BC_flow}.

Equation (\ref{BC_flow}) at $R_1$ corresponds to a no-slip condition
imposed on a wall fixed with  respect to the laboratory frame. This
deserves further discussion. In Eq. (\ref{BC_flow}) the BC is that
$\textbf{u}$ at the surface of the moving particle, i.e.,
$\left.\mathbf{u}\right|_{|\mathbf{r} - \mathbf{R}_p(t)| =  R_+}$
where $\mathbf{R}_p(t) = \int^t_0 \mathbf{V}(t') dt'$ is the position
of the particle center, takes the value $\mathbf{V}(t;\eta) +
\mathbf{v}_s$. As long as $|\mathbf{R}_p(t)| \ll R_+$, concentricity
holds and one obtains the first of the two BCs given by
Eq. (\ref{BC_flow}). If, however, the velocity $\mathbf{V}$ is not
small, we readjust the center of $R_1$ in order to impose
concentricity using the following ``protocol'': the particle is
allowed to move for a short time $\delta t$ with the instantaneous
velocity $\mathbf{V}(t)$ while the spherical shell at $R_1$ is fixed,
after which the spherical shell is displaced by
$\mathbf{V}(t) \delta t$ to its new fixed position in such a way that
it does not perturb significantly the density and flow fields; this
procedure is then repeated. It is not clear to which extent
this latter assumption, which allows us to obtain an analytical
solution, can be realized experimentally. But it is expected that
the results we derive for the present geometrical setup are relevant
also for more general geometries \cite{Zydney_1995}, such as a
particle moving along a channel in \textit{3d}, for which the
constraint that the confinement moves with the particle is not
needed.

We note that in the general case of a non-spherical particle, or of
a particle with non-uniform surface properties (e.g., a spatially
varying effective mobility $b$) the particle can also rotate and the
BC at the particle surface, Eq. (\ref{BC_flow}), should include a
term accounting for a  rigid-body rotation with angular velocity
$\mathbf{\Omega}$. However, this angular velocity turns out to be
identically zero in the case of a spherical particle with an effective
mobility $b$ which is constant on its surface
\cite{Anderson_1989,Morrison_1970}; therefore we completely
disregard it here.

\subsection{Phoretic velocity}

The velocity $\mathbf{V}(t;\eta)$ of the particle is obtained by
requiring that the hydrodynamic force
\begin{equation}
\label{hydro_force}
\mathbf{F} = \iint\limits_{|\mathbf{r}|= R}
\,\mathbf{\hat \Pi}\,\mathbf{e}_r\,dS \,,
\end{equation}
where $\mathbf{e}_r$ is the radial unit vector and $dS$ the surface
area element on the spherical surface $|\mathbf{r}| = R$ (note that
we replaced $R_+$ by $R$, as discussed in the previous subsection),
exerted by the fluid on the composite domain particle plus surface
film vanishes [see the vector identities in Eq. (\ref{vec_ident}),
(i) - (iii) below this equation, and
$(\mathbf{r} \times (\mathbf{\nabla} \times \mathbf{u}))_{\alpha}
= x_{\beta} \partial_{\alpha} u_{\beta}
- x_{\beta} \partial_{\beta} u_{\alpha}$, with summation over
$\beta$]:
\begin{equation}
\label{zero_force}
\iint\limits_{|\mathbf{r}|= R} \,dS \,
\left[-p \,\mathbf{e}_r +
\mu \left(\dfrac{\partial \mathbf{u}}{\partial r}-
\dfrac{\mathbf{u}}{r}\right)
+ \dfrac{\mu}{r}\nabla(\mathbf{r}\mathbf{u})\right]
= 0 \,.
\end{equation}
If, as discussed above, a rotational motion with angular velocity
$\Omega(t;\eta)$ would have to be considered, too, this will be
determined by the additional requirement that the motion is not
only force free but also torque free  \cite{Anderson_1989}. This is
again due to the fact that there are no net forces acting on the
object composed of the particle and its surface film.

The above argument for determining the velocity $V$ has been
discussed in detail by Anderson (see Ref. \onlinecite{Anderson_1989} and
references therein), but it is often overlooked and replaced by
the incorrect argument of a balance between a drag force - i.e., the
integral of the non-uniform osmotic pressure proportional to the
density of solute (see Appendix \ref{app_A}) - exerted on the
particle and a Stokes-like viscous friction from the solvent (see,
e.g., Refs.~\onlinecite{Brady_2008,Paxton_2006,Saidulu_2008}).
It is important to realize that the occurrence of composition
gradients in the solution does not give rise, by itself, to an
osmotic pressure (see also Ref. \onlinecite{Prost}), in contrast to such
an assumption made in Ref. \onlinecite{Brady_2008}. Such gradients will
simply lead to diffusion of the product molecules, while the
pressure in the solvent will adjust to accommodate the spatially
varying chemical potential corresponding to the quasi-stationary
density profile (mildly time dependent due to the overall increase
of the solute number density, in the case of the confined system),
reflecting mechanical equilibrium \cite{Prost}. As a matter of fact,
the origin of this osmotic pressure resides in the interaction
between the particle and the product molecules, i.e., it requires
the explicit consideration of the effective interaction potential
between the particle and the product molecule (for a detailed
illuminating discussion of this point see Refs. \onlinecite{Anderson_1989,Ajdari_2006}).
An intuitive argument that the use of such a ``Stokes-formula'',
which stems from a standard ``drag balanced by viscous friction''
type of reasoning, is incorrect can be formulated as follows. At
distances far from the particle the flow field should look like
that produced by the superposition of a point force $\mathbf{f}$
at the origin, i.e., the center of the particle, which is the
integrated (over the particle surface) effective ``product molecules
on particle'' interaction (the forces $\hat{\mathbf{f}}_\mathcal{D}$
in Fig. \ref{fig3} in Appendix \ref{app_A}), and a distribution of
effective point forces oriented radially (of the particle acting on
the product molecules) in a small shell region around the surface of
the particle (which is the aforementioned surface film), which upon
integration gives exactly $-\textbf{f}$. Note that for repulsive
effective interactions and for an axisymmetric distribution of
product molecules with an increasing density towards the source at
$z = -R$, as in Fig. \ref{fig1}, $\mathbf{f} =
\int_S \hat{\mathbf{f}}_\mathcal{D}\, dS $ is oriented into the
positive $z$-direction; for attractive interactions, $\mathbf{f}$
would be oriented into the negative $z$-direction. This can be seen
as follows. For repulsive interactions, each of the forces
$\hat{\mathbf{f}}_\mathcal{D}$ is oriented radially inwards; since
the density of the product molecules is increasing towards the
source located on the lower hemisphere, the magnitude and the
projection onto the $z$-direction of the force due to the domain
$\cal{D}$ located at any $0 < \theta < \pi/2$ is smaller than the
one in the corresponding domain $\cal{D}$ located at
$\pi - \theta$. Thus the contribution into the positive
$z$-direction from the lower hemisphere will dominate. For
attractive interactions, the argument is simply reversed.
This is in accordance with the following two statements regarding
the characteristics of the far-field (i.e., on length scales over
which the particle plus surface film are seen as point-like)
hydrodynamic flow: (i) In the absence of external body forces such
as gravity or centrifugal forces the motion of the particle plus
its surface film is net force free. (ii) The generated flow
corresponds, within a first approximation, to that produced by a
``force dipole'' [which is the superposition of a distribution of
``force dipoles'' ($\hat{\mathbf{f}}_\mathcal{D}$ acting on the
center and $-\hat{\mathbf{f}}_\mathcal{D}$ on the product molecules
in $\mathcal{D}$) as in Fig. \ref{fig3} in Appendix \ref{app_A}],
or a higher order ``force multipole'', e.g., quadrupole (if it
happens that the net force dipole vanishes, too) \cite{Prost} at
the origin, i.e., the position of the center of the particle,
rather than to the one due to a point force, which would be the
case for an object uniformly dragged against the viscous Stokes
friction (see, e.g., Ref. \onlinecite{Chwang_1975}). For an unbounded
system, these forces translate into a radial decay of the phoretic
flow field proportional to $r^{-2}$ (force dipole) or $r^{-3}$
(force quadrupole), in contrast to the decay proportional to
$r^{-1}$ corresponding to a point force; for a bounded system, the
differences between the flow fields cannot be any longer summarized
by such a simple criterion as different power laws for the radial
decay, but they remain significant nevertheless. These features of
the hydrodynamic flow are discussed in more detail in the
Appendix \ref{app_B} (see also, c.f., Fig. \ref{fig4}).

Using the expansion of the velocity and pressure fields in terms of
the solid harmonics $K_\ell$ (see Appendix \ref{app_B}), the
hydrodynamic force on the particle, defined by
Eq. (\ref{hydro_force}) and expressed as on the left-hand side of
Eq. (\ref{zero_force}), reduces to
\begin{equation}
\label{force_p2}
\mathbf{F} = 4 \pi \tilde p_{-2} \nabla [r P_1(\cos\theta)]\,,
\end{equation}
where $\tilde p_{-2}$ is the coefficient of $K_{-2}$ in the
expansion of the pressure \cite{Happel_book}
[Eqs. (\ref{press_expansion}) and (\ref{solid_harmonics})], and
$P_1$ is the
Legendre polynomial of order one. All other terms vanish since the
corresponding integrals are exactly equal to zero. Thus the
requirement of a vanishing $\mathbf{F}$ implies $\tilde p_{-2} = 0$,
which leads to (see Appendix \ref{app_B})
\begin{equation}
\label{velocity}
V(t) = \chi_1(\eta) \dfrac{b}{R}
\int\limits_{0}^{\pi} \,d\theta \,\sin\theta\, \cos\theta \,\,
\rho(R,\theta, t;\eta)
\end{equation}
where
\begin{equation}
\label{def_chi1}
\chi_1(\eta) = 1 - \dfrac{5}{2}\dfrac{\eta^2-1}{\eta^5-1}\,.
\end{equation}
The structure of the expression on the right-hand side (rhs) of
Eq. (\ref{velocity}) clarifies the meaning of the factor
$\chi_1(\eta)$, which varies between zero, at $\eta \to 1$, and one,
at $\eta \to \infty$ (see Fig. \ref{fig2}). Without this factor,
one has on the rhs the phoretic velocity in the unbounded space due
to a source which generates a composition profile $\rho_{\infty}$
such that $\rho_{\infty}(R,\theta,t) \equiv \rho(R,\theta, t;\eta)$
at all times $t$, as derived in Ref. \onlinecite{Golestanian_2005}
[with $\Lambda \to -\lambda^2$ entering into $b$; see the text
around Eq. (\ref{lambda_def})]. Thus $\chi_1(\eta)$ is a
``hydrodynamic wall-correction'' factor which quantifies and
summarizes the effects solely due to the confinement induced changes
in the characteristics of the solvent flow. Note that
$\chi_1(\eta) < 1$ means that the hydrodynamic effects due to
confinement will tend to decrease the velocity from the value
corresponding to an unbounded system.

\subsection{Density profile of the product molecules and
diffusiophoretic velocity}

According to Eq. (\ref{velocity}), knowledge of the density
$\rho(|\mathbf{r}| = R_+,t;\eta)$ of product molecules completely
determines the velocity $\mathbf{V}$ of the particle as a function
of $\eta < \infty$ for confined systems and thus allows one to
quantify the effects of confinement on the resulting phoretic
motion.

Within the assumptions that the diffusion of product molecules is
fast compared with the convection by the solvent flow, i.e., in the
limit of small Peclet numbers, and that the product distribution
$\rho(\mathbf{r},t)$ is undisturbed by the flow, i.e., neglecting
any so-called polarization effects of the surface film \cite{Anderson_1989},
the time evolution of the number density $\rho(\mathbf{r},t)$ of
product molecules around the moving particle is governed, in the
co-moving frame, by the diffusion equation
\begin{equation}
\label{diff_eq}
\partial_t \rho = D \nabla^2 \rho +
B(t) \delta(\mathbf{r}-{\mathbf{r}}_s),~~R_+ < |\mathbf{r}| < R_1\,.
\end{equation}
This equation is to be solved subject to the initial condition
(IC) of zero density of product molecules and to the boundary
conditions of zero normal current at the confining wall (assuming
that the latter is co-moved, without perturbing the solution, such
that at all times it remains concentric with the particle) and
at the outer edge $r = R_+$ of the surface film. Actually, the
latter condition should be imposed at the surface of the particle,
but we shall consider only the case of a thin surface film so that
due to its small volume the transport of solute to the surface
film is negligible compared to the transport in the outer region
and the zero normal current condition at $R_+$ is a good
approximation. Hence, the IC and BCs are:
\begin{equation}
\label{BC_IC}
\rho(\mathbf{r},0) = 0,~~\left.\left(
\frac{\partial \rho(\mathbf{r},t)}{\partial r}
\right)\right|_{|\mathbf{r}|= R_+, R_1} = 0.
\end{equation}

We remark that the linearity of Eqs. (\ref{diff_eq}) and
(\ref{velocity}) with respect to $\rho$ imply that by superposition
the solution of the present problem can be extended to the case of
an arbitrary spatial arrangement of several catalytic sites on the
particle surface. This will allow one to find an optimal decoration
for providing stability against particle rotations (caused by thermal
noise), which can otherwise spoil the unidirectional motion (see
Ref.~\onlinecite{Golestanian_2007} for such examples of designed
optimal surface distributions).

Equation (\ref{diff_eq}) subject to the IC and BC conditions given
by Eq. (\ref{BC_IC}) is solved by using the Laplace transform.
Similarly to the approximation employed in
Sec. \ref{subsec_phor_flow}, from this point further in the
calculations we replace everywhere $R_+$ by $R$ because the
difference $\delta$ (the thickness of the surface film) between the
two is negligible on the macroscopic length scales characterizing
the transport in the outer region. The solution is obtained as a
series in products of Legendre polynomials and modified spherical
Bessel functions of the first and third kind, respectively; the
details are provided in Appendix \ref{app_C}. Focusing on the
particular case in which after its start the activity of the
catalytic site is time independent, i.e., $B(t)$ is constant:
$B(t) = H(t)/\tau_f$, where $H(t)$ is the Heaviside step function
and $\tau_f$ is the average production time for the creation of a
product molecule. The Laplace transform of
the integral in Eq. (\ref{velocity}) is computed by using the
Laplace transformed density $\rho(\mathbf{r},\zeta;\eta)$ and the
fact that $\cos \theta = P_1(\cos \theta)$. Formally inverting the
Laplace transform, one obtains the diffusiophoretic velocity as a
function of time and confinement:
\begin{equation}
\label{velocity_eta}
V\left(s=\dfrac{D t}{R^2};\eta \right)
= \dfrac{b \chi_1(\eta)}{\pi^2 R^2 D \tau_f}
\left({\cal L}^{-1}[\bar \Phi (\xi;\eta)]\right)|_s\,,
\end{equation}
where
\begin{eqnarray}
\label{Phi_def}
\bar \Phi (\xi;\eta) &\equiv& \frac{i_{3/2}(\sqrt{\xi})}{\sqrt{\xi}} 
\nonumber\\
&\times& \left[ {\hat\alpha}_1 \,i_{3/2}(\sqrt{\xi})
+
({\hat\beta}_1-1) \,k_{3/2}(\sqrt{\xi})
\right]
\,,\hspace*{.2in}
\end{eqnarray}
$i_{\ell+1/2}(z)
= \sqrt{\pi/(2 z)} I_{\ell+1/2}(z)$ and $k_{\ell+1/2}(z) =
\sqrt{\pi/(2 z)} K_{\ell+1/2}(z)$ are modified spherical Bessel
functions of the first and third kind, respectively
\cite{Abramowitz_book}, and the dimensionless coefficients
${\hat\alpha}_1 (\sqrt{\xi},\eta)$ and
${\hat\beta}_1 (\sqrt{\xi},\eta)$ are fixed by imposing the boundary
conditions [see Eq. (\ref{alpa_beta_3d_eqs}) and Appendix \ref{app_C}
with $\xi = \zeta R^2/D$ so that $\xi$, $\bar \Phi$, and
${\cal L}^{-1}[\bar \Phi]$ are dimensionless].

\section{Discussion}
\label{discussion}

\begin{figure}[!htb]
\includegraphics[width=.9\linewidth]{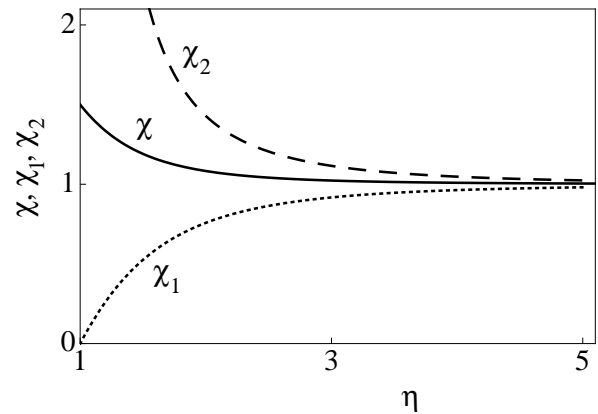}
\caption{
The long-time asymptotic velocity
$V^{(\infty)}(\eta)/V_0 \equiv \chi(\eta) > 1$ (solid line)
as a function of the confinement $\eta$ (see Fig. \ref{fig1})
and the hydrodynamic and diffusion ``wall-correction'' factors
$\chi_1(\eta) < 1$ (dotted line) and $\chi_2(\eta) > 1$
(dashed line).
}
\label{fig2}
\end{figure}

The complicated structure of the function $\bar \Phi (\zeta;\eta)$
makes it rather laborious to carry out the inverse Laplace
transform, so that the full time dependence of the velocity $V(t)$
cannot be derived easily. However, the asymptotic
($s = t D/R^2 \gg 1$) value of the velocity
$V^{(\infty)}(\eta):= {\displaystyle \lim_{s \to \infty}} V(s;\eta)$
can be straightforwardly derived by using the inversion formula for
the Laplace transform \cite{Carslaw_book} by noticing that
$\bar \Phi (\xi;\eta)$ has a simple pole at $\xi = 0$,
which determines the asymptotic value of the velocity as the residue
of $\bar \Phi (\xi;\eta)$ at $\xi = 0$ \cite{Carslaw_book}.
The existence of such a constant asymptotic value, which at first
glance seems to be in conflict with the fact that the density
$\rho(\mathbf{r};t)$ does not reach a steady state (since we consider
a closed system with a source continuously producing non-interacting
point-like particles), is due to the fact that the phoretic motion is
determined solely by the gradient of the number density of the
product molecules along the surface of the particle. At long times,
the total density of product molecules is large and any redistribution
(which would lead to a change of the density gradient) proceeds on
very slow time scales. (Ultimately the density of product molecules
becomes so high that they no longer behave as an ideal gas and the
latter assumption breaks down.) This yields
\begin{equation}
\label{Vasymp}
V^{(\infty)}(\eta)
= \dfrac{b \chi_1(\eta)}{\pi^2 R^2 D \tau_f}
\mathrm{Res}_{\xi=0}\,[\bar \Phi (\xi;\eta)] =
V_0 \chi(\eta)\,,
\end{equation}
where
$V_0 = - \dfrac{b}{4 \pi R^2 D \tau_f} = V^{(\infty)}(\eta\to\infty)$
is the asymptotic velocity in the case of an unbounded system and
[see Eq. (\ref{def_chi1})]
\begin{equation}
\label{chi}
\chi_(\eta)=\chi_1(\eta)\frac{\eta^{3}+2}{\eta^{3}-1}
:=\chi_1(\eta)\chi_2(\eta)\,,~\forall \, \eta > 1 \,.
\end{equation}
Thus $\chi_2(\eta) = (\eta^{3}+2)/(\eta^{3}-1)$ is a ``diffusion
wall-correction'' factor, and $\chi(\eta)$ quantifies the combined
wall effects, which in the present case factorize into a
hydrodynamic and a diffusion contribution. These two contributions
oppose each other such that the confinement in hydrodynamics
decreases the particle velocity ($\chi_1 < 1$) whereas the
confinement of the diffusion enhances it ($\chi_2 > 1)$ (see
Fig. \ref{fig2}); the latter dominates so that $\chi > 1$ and
there is an overall enhancement of the velocity
(see Fig. \ref{fig2}).

Assuming the Stokes-Einstein relation for the diffusion coefficient
of the product molecules in the solvent (see Sec. \ref{intro}) and
using the expression Eq. (\ref{b_def}) for the mobility $b$, the
velocity $V_0$  [Eq. (\ref{Vasymp})] can be rewritten as
\begin{equation}
\label{V0_simple}
V_0 =  - \dfrac{3}{4} \, \dfrac{\Lambda}{|\Lambda|}  \,
\left(\frac{\lambda}{R}\right)^2 \dfrac{a}{\tau_f} \,.
\end{equation}
Surprisingly, this expression has no explicit dependence on the
viscosity of the solvent or on the temperature. In fact, the
dependence on viscosity drops out upon using the Stokes-Einstein
relation. In turn, the dependence on $T$ is hidden in the temperature
dependence of the (effective) interaction between the particle and
the product molecules which determines the thickness of the surface
film, i.e., the length scale $\lambda$ [see Eq. (\ref{lambda_def})].
In the case of only steric repulsion between the particle and the
product molecules, for a particle of radius $R = 1~\mathrm{\mu m}$
[$100 \mathrm{nm}$], a product molecule with diameter
$a = 1.0~\mathrm{nm}$, $\lambda = 1.0~\mathrm{nm}$ and a reaction
rate (following Ref. \onlinecite{Golestanian_2005}) of
$1/\tau_f = 25~\mathrm{kHz}$, one finds the velocity
$V_0 \simeq 10$ pm/s [1 nm/s]; consequently $V^{(\infty)}(\eta)$
is of the same order of magnitude. Note that these values are
much smaller than the estimate $V_0 = \mathcal{O}(\mathrm{\mu m/s})$
of Ref. \onlinecite{Golestanian_2005}, which is based on a significantly
larger value $\lambda^2 \simeq 10^{-15}~\mathrm{m}^2$ than the one
$\lambda^2 \simeq 10^{-18}~\mathrm{m}^2$ corresponding to the
reasonable estimate $a = 1$ nm for the diameter of the product
molecule. However, this estimate corresponds to the case of a single
catalytic site. One can argue that several such single reaction
sites distributed closely around $\mathbf{r}_s$ may lead to a
significant increase of the velocity (intuitively, by a factor equal
to the surface density of such reaction sites). Thus the resulting
velocities may eventually be closer to the experimentally reported
values \cite{Paxton_2006,Howse_2007} $V_0 =
\mathcal{O}(\mathrm{\mu m/s})$ for objects with a $\mu$m$^2$ area
covered by catalyst if the surface density of reaction sites is of
the order of nm$^{-2}$. Moreover, if the effective interaction
between the particle and the product molecules is attractive, the
ratio $(\lambda/a)^2$ can be very large \cite{Anderson_1989}.
Another possibility leading to a significant increase is that of
a classical hydrodynamic slip boundary condition with a large
slip-length at the surface of the particle for the flow within
the surface film, assuming that (see Appendix \ref{app_A}) classical
hydrodynamics provides an accurate description for the solvent
flow even at such small scales \cite{Ajdari_2006}. (According to the
text following Eq. (\ref{g_def}), for the derivation of the effective
mobility $b$ in the phoretic slip-velocity [Eqs. (\ref{slip_vel}),
(\ref{b_def}), and (\ref{slip_vel_hatx})] we have used a no-slip
boundary condition for the flow in the surface film at the particle
surface. This condition can be generalized to a slip condition
characterized by a slip length.) Such hydrodynamic slip may be, e.g.,
due to roughness of the particle surface, and its effect on the
phoretic motion can be intuitively understood as follows. The slip
over the surface of the particle facilitates the solvent flow within
the surface film as compared to the case of zero slip (a
``lubrication'' effect), and thus for the same pressure gradient
along the surface as in the case of zero slip the flow velocity in
the surface film will be increased for nonzero slip-lengths. The
phoretic slip $\mathbf{v_s}$,  which is the flow velocity at the
outer boundary of the surface film, is thus also increased compared
to its value in the case studied here so far that there is a no-slip
condition at the surface of the particle; this leads to an increase
(by up to orders of magnitude \cite{Ajdari_2006}) of the phoretic
velocity of the particle.

The velocity $V^{(\infty)}(\eta)$ of the particle can be positive
or negative, depending on the details of the interaction potential.
This means that the particle can travel either following the gradient
of solute density, or against it. According to Eqs. (\ref{Vasymp})
and (\ref{V0_simple}) as well as the definition of the
effective mobility $b$, the sign of the phoretic velocity is opposite
to the sign of the parameter $b$ (or, equivalently, that of the
parameter $\Lambda$), which in turn depends, inter alia, on the
attractive or repulsive character of the effective interactions
between the particle and the solute molecules [see
Eq. (\ref{lambda_def})]. Equations (\ref{slip_vel}) and
(\ref{surf_grad}) show that the sign of the phoretic slip velocity,
i.e., of the polar component $\mathbf{v}_s \cdot \mathbf{e}_\theta$
is also opposite to the one of $b$, because for our system
$\partial_\theta \rho$ is positive (see Fig. \ref{fig1}). In the case
of hard core interactions only one has $\Lambda < 0$ and thus
$b < 0$; this implies $V^{(\infty)}(\eta) > 0$, i.e., the particle
moves in the direction of $\mathbf{e}_z$ and thus \textit{away} from
the catalytic site, while the slip velocity points in the same
direction as $\mathbf{e}_\theta$, which has a negative $z$-component,
so that the flow around the particle is also oriented towards the
catalytic site \cite{Anderson_1989}. The confinement does not change
the direction of $\mathbf{V}$, which for an unbounded system is, as
expected, in agreement with Ref. \onlinecite{Golestanian_2005}. A simple
intuitive explanation for the direction of the motion of the particle
follows from the observation that there is no net force acting on the
composite consisting of the particle plus the surface film, so that
there is conservation of momentum. Because the solvent flow around
the particle is in the negative $z$-direction [see
Fig. \ref{fig4}(a)], the particle should move in the positive
$z$-direction. Note that while this simple argument is clear in the
unbounded case, for the confined system it breaks down because the
formation of vortices [see Fig. \ref{fig4}(c)] implies that there are
spatial regions where the flow is in the positive $z$-direction. Thus
a quantitative analysis is required.

If $b > 0$, on the contrary, $V^{(\infty)}(\eta) < 0$, which  means
that the particle will move \textit{towards} the source of the product
molecules. This is in contrast to the results presented in Ref. \onlinecite{Brady_2008}, which predict motion \textit{always away}
from the source. As discussed before the reason for this discrepancy
is the ignorance in Ref. \onlinecite{Brady_2008} of the mediating role of
the solvent and its assumption of a mapping of the non-uniform
density of the solute onto a non-uniform ``osmotic pressure'' acting
on the particle (i.e., assuming that more product molecules impinge
on the particle from the side with higher density), rather than
mapping onto gradients in the solution pressure along the surface
of the particle within its surface film.  This opposing
directionality in the case $b > 0$ thus provides a simple criterion
for an experimental discrimination between the predictions of the
two proposals.

Spatial confinement leads to an overall enhancement of the phoretic
motion, because $|V_0| < |V^{(\infty)}(\eta)|$ for all finite $\eta$.
As shown in Fig.~\ref{fig2}, even at moderate values of $\eta$ the
effect of the confinement is important,  e.g., at $\eta \simeq 2$
there is a ca. $8\%$ increase in the velocity compared with the
unbounded case, and at $\eta \simeq 1.5$ the increase reaches ca.
$25\%$. For $\eta \to 1$ the velocity stays finite. This is, of
course, an artifact which stems from the assumption that the product
molecules are point-like. Actually, there is a lower cut-off $R_1^c$,
where $R_1^c - R$ is of the order of the hard core diameter $a$ of
the product molecules, below which this assumption breaks down and
Eq. (\ref{Vasymp}) is no longer valid. However, for $a \ll R$, which
is a reasonable assumption, one has $\eta_c = 1 + a/R \gtrsim 1$,
and thus the steep increase of the velocity near $\eta = 1$ is
physically relevant. This is a clear counterexample for the statement
in the conclusions of Ref. \onlinecite{Golestanian_2008} that ``the mobility
of such swimmers will be hindered by channel boundaries''.

Note that the velocity remains non-zero and finite for all values
of $\eta$, thus the two opposing effects described by the factors
$\chi_1$ and $\chi_2$ are of similar magnitude. This can be
understood qualitatively from the behavior of the velocity and
density fields in the unbounded system: the hydrodynamic flow
velocity decays as $r^{-3}$, where $r$ is the distance from the
center of the particle \cite{Anderson_1989}, and the gradients in
the number density of product particles (for phoresis the gradients
are relevant, not the density in itself) are are also varying as \cite{Golestanian_2005} $r^{-3}$ . Accordingly the confinement
becomes relevant for both fields at similar length scales and with
a similar power-law behavior. Therefore one can expect that the
confinement has an equally strong influence on the hydrodynamics
and the diffusion.

\section{Summary and conclusions}
\label{summary}

In summary, we have discussed the effect on the phoretic velocity
of a self-propelled particle due to a confining wall for the solvent
and for the reaction product emerging from a catalytic site on the
surface of the dissolved particle.

The analysis is based on considering the present model in the context
of the standard theory of phoresis. We have critically analyzed the
assumptions involved in such an approach. Some of them are already
present in the classical theory, others arise as a result of mapping
the description of such ``active'' surface particles onto the framework
of a theory developed to describe the case of inert particles immersed
in a pre-defined, externally controlled concentration gradient. For
example, within the standard theory of phoresis the perturbation of
the steady-state, externally controlled concentration gradient is
induced by the interaction with the surface of the particle
\textit{upon} immersion. While this is an acceptable concept in that
context, it is not clear how one can justify this if the concentration
gradient develops as a function of time with the particle already
present, as it is the case for a self-propelled object as the one we
have discussed. It is beyond the scope of the present work trying to
improve the general theory, but we consider a clear understanding of
its limitations and of its possible shortcomings as a crucial step
for both avoiding confusions such as those involving the application
of a Stokes-force argument and for future theoretical developments.
The main conclusion from this part of our work is that the development
of a microscopic model for the dynamics in the surface film, eventually
along the lines of Ref. \onlinecite{Kapral_2007} which treats the solvent
and the solute molecules on equal footing, seems to be very important.

Within the confines of the standard theory of phoresis, we have
shown that the presence of a confining wall for the solvent and for
the reaction product emerging from a catalytic site on the surface
of a dissolved particle leads to a significant increase of the
velocity of the self-propelled particle.
This results from two competing effects: an increase of the solute
density gradients along the surface of the particle and a
simultaneous increase of the hydrodynamic viscous friction. The
former one dominates. If only  steric repulsion between the particle
and the product molecules is present, the absolute value of the
velocity is expected to remain, in general, rather small. A direct
experimental realization of the co-moving geometry considered here
seems to be difficult. But the results which we have derived are
expected to be applicable (at least qualitatively) for more general
geometries (see Ref. \onlinecite{Zydney_1995}), such as the motion of
spherical particles along cylindrical tubes. In this sense, an
experimental test of our results may be possible.

Further extensions may focus on new phenomena emerging from more
complicated geometries, as well as from relaxing the assumption of
no interaction between the product molecules. By taking into account
the actual finite size of the product molecules and by considering
generic chemical reactions, in which the reaction products are indeed
different from the solvent molecules, it is expected that the
production of such solute particles is accompanied by a non-uniform
depletion of the solvent around the particle. Thus in general the
motion of the particle will be determined by the gradients in both
the solvent and the solute (see also Ref. \onlinecite{Kapral_2007}).
Moreover, such a depletion zone would also lead to a decrease of the
production rate of the catalytic site and consequently, to a
reduction of the density gradients. Accordingly one might expect that
there are optimal values for the reaction rates for which the
phoretic velocity is maximal.

The interaction between the product molecules may play a significant
role, and this deserves further discussion. If the density of the
product particles is not low (which is reasonable to expect at least
near the catalytic reaction sites, for fast reactions, and for slow diffusion of product particles) and the (solvent-mediated)
solute-solute interactions have attractive or repulsive components
longer ranged than the hard-core interaction discussed above, several
other effects have to be carefully considered, especially if they
are as important as the particle-solute interactions. In this case,
the distribution of the solute particles in the direction normal to
the surface of the big particle is no longer given by
Eq. (\ref{Boltzmann}), but
it is determined  both by the particle-solute interactions and by the contributions of the solute-solute interactions, the latter depending
on the whole distribution of solute particles around the big particle.
Thus it becomes a complicated non-local problem which cannot be
easily addressed analytically. Within a mean-field approximation,
one may still assume that Eq. (\ref{Boltzmann}) holds if the potential
is modified
to include an averaged, effective solute-solute interaction. It is
evident that even such a simplistic approach will lead to a different expression for the phoretic-slip velocity, and thus one can reasonably expect qualitative differences compared to the predictions of the
classic theory of phoresis. Moreover, the solute-solute interactions
will lead to a dynamics which differs from simple diffusion and is determined, roughly speaking, by the density- and
interaction-dependent collective diffusion coefficient. For example,
attractive solute-solute interactions will lead to a tendency of ``clustering'' and thus will hinder the relaxation of the solute gradients, which intuitively would lead to an increase in the
velocity of the particle, while repulsive interactions will help to dissipate the particle gradients and, intuitively, would lead to a
decrease in the velocity. However, such intuitive arguments for
changes in the velocity have to be carefully considered because,
e.g., it is not clear if in the presence of significant
solute-solute interactions the assumption of an ultrathin
``unpolarized'' surface film still holds.

\acknowledgments

The authors thank Prof. U. Seifert for very useful remarks concerning
the role of the solvent and the phoretic hydrodynamic flow. M.N.P.
gratefully acknowledges very fruitful discussions with A. Gambassi,
L. Harnau, and M. Tasinkevych. G.O. acknowledges partial financial
support by Agence Nationale de la Recherche (ANR) under the grant
``DYOPTRI - Dynamique et Optimisation des Processus de Transport
Intermittents''. M.N.P. and G.O acknowledge the hospitality of the
Max-Planck Institute f\"ur Metallforschung (MPI-MF) in Stuttgart, as
well as partial financial support by the MPI-MF.

\appendix

\section{\normalsize Calculation of the phoretic slip velocity}
\label{app_A}
The following considerations are connected to the corresponding
ones in Ref. \onlinecite{Anderson_1989} and put them into the context
of the microscopic model discussed in Sect. \ref{Model}.

\begin{figure}[!htb]
\includegraphics[width=.9\linewidth]{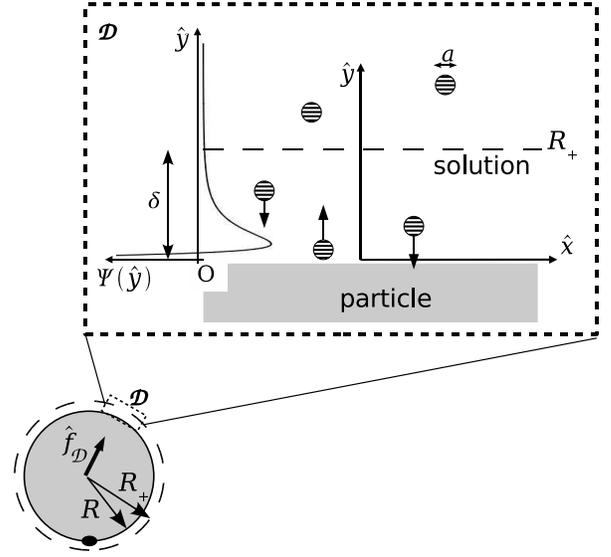}
\caption{
\label{fig3}
Schematic expanded view of a domain $\mathcal D$ of the contact
region between the particle and the solution, far from the reaction
site, approximated locally by a planar geometry. The effective
interaction between the reaction products, shown as horizontally
hatched circles, and the particle is described by an effective
potential $\Psi(\hat y)$; $\hat y$ is the normal distance from the
surface of the particle, also schematically depicted. The arrows at
the product molecules indicate the forces $-\nabla \Psi(\hat y)$
acting on them, while the arrow at the particle center shows the
force ${\hat f}_\mathcal{D}$ on it, due to the same interaction with
the product molecules, which is equal in magnitude and opposite to
the sum of $-\nabla \Psi(\hat y)$ over the product molecules
positions in $\mathcal D$. Here the case is shown that the product
molecules in $\mathcal D$ exert a net attractive force on the
particle. The ``outer edge'' $R_+$ of the surface film beyond which
the effective interactions between the particle and the product
molecules are negligible is indicated by the horizontal dashed line.
The solvent, within which the product molecules move, is taken to
be a homogeneous background (not shown). Actually,
$\delta = R_+ - R$ (of the order of the product molecules diameter
$a$) is considered to be smaller than indicated here.
}
\end{figure}

Due to the azimuthal symmetry of the system, the flow field
$\mathbf{u}(\mathbf{r})$ of the solvent has non-zero components
only along the radial and polar directions. According to the
considerations in Sect. \ref{Model}, the dissolved product
molecules (solute) of diameter $a$ are exposed to an effective
interaction with the particle within a surface film of thickness
$\delta \sim a$. Accordingly, the solute molecules do not interact
with the particle beyond the radial distance $R_+ = R + \delta$
(measured from the center O of the particle). Since the particle
radius $R$ is much larger than $a$ so that $\delta/R \ll 1$, this
surface film of thickness $\delta$ can be approximated to be
locally planar (see Fig. \ref{fig3}).
In a small domain $\mathcal D$ centered at $\mathbf r =
(R,\theta,\phi)$ (Fig. \ref{fig3}), the local coordinate system
(\textasciicircum) is chosen such that $\hat x$ is along the polar
direction $\mathbf{e}_\theta$ while $\hat y$ is along the radial
(normal) direction $\mathbf{e}_r$.

Within the planar film approximation, which is reliable because the
film thickness $\delta$ is taken to be much smaller than the particle
radius $R$ ($\delta \ll R$), and for slow and smooth variations of
the solute density along the surface of the particle (i.e., the
solute density is assumed to vary over length scales of the order of
the particle radius $R$), the flow of the solvent relative to the
particle surface is $\hat{\mathbf{u}} (\hat x, \hat y) \simeq
({\hat u}_{\hat x} (\hat x,\hat y), 0)$ (with the meaning that the
ratio ${\hat u}_{\hat y}/{\hat u}_{\hat x}$ is of the order
${\cal O}(\delta/R) \ll 1$) \cite{Anderson_1989}. This can be
intuitively understood starting from the incompressibility condition
$\hat \nabla \hat{\mathbf{u}} = 0$, where $\hat \nabla$ indicates
that the derivatives are taken with respect to the local coordinates.
The component ${\hat u}_{\hat x}$ is of the order of the
slip velocity $|\mathbf{v}_s|$, which is expected to be
proportional to the gradient along the surface, i.e., along the
$\hat x$ direction, of the solute density $\rho$ evaluated at $R_+$.
Noting that $d \hat x = R d \theta$, one obtains
$\partial_{\hat x} {\hat u}_{\hat x} \sim
(1/R^2) \,\partial_\theta^2 \rho$. On the other hand, the component
${\hat u}_{\hat y}$ has to rise steeply within the surface film of
thickness $\delta$ from the value zero at the surface of the particle
to the value corresponding to the radial component of the outer flow,
which is expected to be also of the order of $|\mathbf{v}_s|$.
Therefore $\partial_{\hat y} {\hat u}_{\hat y} \sim
|\mathbf{v}_s|/\delta \sim [1/(R \delta)] \,\partial_\theta \rho$.
Since the derivatives of the solute density profiles with respect to
$\theta$ are expected to not depend on the surface film thickness
$\delta$ and to be finite (due to assumed smooth, slow variations
over length scales of the order of the particle radius), it follows
that $\partial_{\hat y} {\hat u}_{\hat y}/
\partial_{\hat x} {\hat u}_{\hat x} \sim R/\delta$ so that
$|\partial_{\hat y} {\hat u}_{\hat y}| \gg
|\partial_{\hat x} {\hat u}_{\hat x}|$. Thus the incompressibility
condition $\partial_{\hat x} {\hat u}_{\hat x} +
\partial_{\hat y} {\hat u}_{\hat y} = 0$ implies that the leading
contribution $\sim 1/(R \delta)$ must vanish separately because it
cannot be canceled by subleading contributions $\sim 1/R^2$.
Therefore one has $\partial_{\hat y} {\hat u}_{\hat y} \simeq 0$ (up
to terms of relative order ${\cal O}(\delta/R)$). This implies that
${\hat u}_{\hat y}$ does not depend on $\hat y$; because
${\hat u}_{\hat y} = 0$ on the surface of the particle
(independently of $\hat x$), due to the surface being impermeable,
it follows that ${\hat u}_{\hat y} \simeq 0$ (in the sense of
correction terms $\sim \delta/R$) everywhere in ${\cal D}$.

As indicated in Fig. \ref{fig3}, the effective interaction of the
product molecules (solute) with the particle, characterized by the
effective interaction potential $\Psi(\hat y)$ (in the limit of low
solute density many-body effects can be ignored so that $\Psi$ is
independent of $\hat x$), gives rise to a force $ - \nabla \Psi$
acting on the solute molecules. Assuming that the solvent can be
approximated as a continuum on the length scale of the solute, the
no-slip condition at the surface of the \textit{solute} molecules
implies that a body force density $ - \rho \nabla \Psi$ is
transmitted to the solvent. (Note that, as discussed also in the
main text, the approximation of the solvent as a continuum breaks
down if the solute molecules have a size similar to that of the
solvent molecules, as in the case of the experiments discussed in
Refs. \onlinecite{Paxton_2005,Howse_2007}; in such a case this
transmission of the body force has to be considered as an assumption
to be verified \textit{a posteriori}.)
Evidently, such forces are present only within the surface film,
where the effective interaction potential is non-zero. Within these
assumptions, the Stokes equations for the flow
$\hat{\mathbf{u}} (\hat x, \hat y) \simeq
({\hat u}_{\hat x} (\hat x,\hat y), 0)$
within the surface film take on the form:
\begin{equation}
\label{Stokes_complete}
\mu {\hat \nabla}^2 \hat{\mathbf{u}} =
\hat \nabla p + \rho \hat \nabla \Psi\,.
\end{equation}
The equation corresponding to the $\hat x$ component of
Eq. (\ref{Stokes_complete}) can be further simplified by noting
(with an argument similar to the one used in the paragraph
above) that the partial derivative
$|\partial_{\hat x}^2 {\hat u}_{\hat x}|$
is much smaller than $|\partial_{\hat y}^2 {\hat u}_{\hat x}|$ (by a
factor of the order ${\cal O}[(\delta/R)^2$]). Thus for the $\hat x$
component Eq. (\ref{Stokes_complete}) leads to
\begin{subequations}
\label{inner_Stokes}
\begin{equation}
\label{x_inner_Stokes}
\mu \partial^2_{\hat y} {\hat u}_{\hat x} =
\partial_{\hat x} p\,,
\end{equation}
whereas the $\hat y$ component yields
\begin{equation}
\label{y_inner_Stokes}
\partial_{\hat y} p = - \rho(\hat x, \hat y)
\partial_{\hat y} \Psi\,.
\end{equation}
\end{subequations}

We further assume that within the very thin surface film the
relaxation of the solute density profile along the direction normal
to the surface of the self-propelled particle towards a steady-state
(zero net current, i.e., the diffusion current due to gradients in
$\rho$ along the $\hat y$ direction is balanced by the convective
current generated by the force-field $-\nabla \Psi$) is fast compared
to the diffusional relaxation time of its gradient along the surface
\cite{Anderson_1989,Ajdari_2006}. In this case, \textit{within} the
surface film the number density of the solute, assumed to behave
like an ideal gas, is given by the Boltzmann distribution
corresponding to the effective interaction $\Psi (\hat y)$:
\begin{equation}
\label{Boltzmann}
\rho(\hat x, \hat y) \simeq
\rho(\hat x, R_+) e^{-\beta \Psi(\hat y)}\,.
\end{equation}
(The prefactor $\rho(\hat x, R_+)$ reflects the fact that
$\Psi (\hat y \to \delta)$ becomes negligibly small; we emphasize
that Eq. (\ref{Boltzmann}) applies \textit {only within} the
surface film.) We note that this assumption can break down,
especially in the immediate vicinity of the reaction site, if the
reaction rate is high. In such a case one has to explicitly consider
the coupled equations of mass transport for the solvent and solute
within the surface film (see Ref. \onlinecite{Kapral_2007}); as already
mentioned before, we consider here only situations in which
Eq. (\ref{Boltzmann}) holds.

Combining Eqs. (\ref{y_inner_Stokes}) and (\ref{Boltzmann}),
integrating with respect to $\hat y$, and using that the inner
and outer solutions for the hydrodynamic flow should match
smoothly at $R_+$ so that at $R_+$ the pressure reaches its
``outer'' solution value $p_\mathrm{out}(\hat x, \hat y = R_+)$, one
obtains the following expression for the pressure field within the
surface film:
\begin{equation}
\label{inner_pressure}
p(\hat x,\hat y) = p_{\mathrm{out}} (\hat x, R_+)
+ k_B T [\rho(\hat x, \hat y)-\rho(\hat x,R_+)]\,.
\end{equation}
This is the so-called ``osmotic equilibrium'' condition (because on
the right hand side the term after the plus sign has the form known
as ``osmotic pressure'') along the surface normal, which varies
along the surface of the particle due to the gradient in the solute
density \cite{Anderson_1989,Ajdari_2006}. Note that
Eq. (\ref{inner_pressure}) also applies \textit{only within} the
surface film.

By: (i) combining Eqs. (\ref{x_inner_Stokes}), (\ref{Boltzmann}),
and (\ref{inner_pressure}), (ii) noting that $\partial_{\hat x}
p_{\mathrm{out}} (\hat x, R_+)$ can be neglected because
$\partial_{\hat x} p_{\mathrm{out}} (\hat x, R_+) \sim
\mu |\mathbf{v}_s|/R^2$ (this is because the outer solution
satisfies the force-free ($\Psi = 0$) version of
Eq. (\ref{Stokes_complete}); by evaluating it at $R_+$ and along
the surface of the particle and by noting that the slip velocity
varies over the macroscopic length scale $R$, the above conclusion
follows) while with
$|{\hat u}_{\hat x}(\hat y = R_+)| = |\mathbf{v}_s|$ and
${\hat u}_{\hat x}(\hat y = 0) = 0$ one finds from
Eq. (\ref{x_inner_Stokes}) that $\partial_{\hat x} p(\hat x, R_+)
\sim \mu |\mathbf{v}_s|/\delta^2$, (iii) introducing
\begin{equation}
\label{h_def}
h(\hat x, \hat y):=
\dfrac{k_B T}{\mu} \dfrac{d \rho(\hat x, R_+)}{d \hat x}
\left(e^{-\beta \Psi(\hat y)}-1\right)\,,\nonumber
\end{equation}
(iv) integrating once, and (v) using
$\partial_{\hat y} {\hat u}_{\hat x}|_{R_+} = 0$ (which holds because
the inhomogeneity causing the flow vanishes at the outer edge of the
surface film), one obtains
\begin{equation}
\label{g_def}
\partial_{\hat y} {\hat u}_{\hat x} =
- \int\limits_{\hat y}^{R_+} \,du\, h(\hat x,u) =:
- g(\hat x,\hat y)\,.
\end{equation}
Integrating by parts, using the boundary condition
${\hat u}_x(\hat y = 0) = 0$ of no-slip at the particle surface,
and noting that $u \, g(\hat x,u) \to 0$ for $u \to 0$ (because
$h(\hat x,u)$ is bounded with respect to $u$, so that $g(\hat x,u)$
is finite for all $u$) and $g(\hat x,R_+) = 0$ [by definition of
$g(\hat x,\hat y)$], one obtains the phoretic
slip-velocity $\mathbf{v}_s =
{\hat u}_x(\hat x = R_+ \theta,R_+) \, \mathbf{e}_\theta$ as
\begin{equation}
\mathbf{v}_s (\theta) = - \mathbf{e}_\theta
\int\limits_{0}^{R_+} \,du\, u \,h(\hat x = R_+\theta, u)\,.
\label{slip_vel_hatx}
\end{equation}
Since $h(\hat x,u)$ decays rapidly for $u > R_+$ the integral
in Eq. (\ref{slip_vel_hatx}) can be extended to infinity without
causing a significant error:
\begin{eqnarray}
\mathbf{v}_s (\theta) &\simeq& 
- \mathbf{e}_\theta\,
\dfrac{k_B T}{\mu} \dfrac{d \rho(\hat x , R_+)}{d \hat x}
\int\limits_{0}^{\infty} \,du\, u \,[e^{-\beta \Psi(u)}-1]
\hspace*{.1in}\nonumber\\
&=& - \mathbf{e}_\theta \, b \,
\dfrac{d \rho(\hat x = R_+ \theta, R_+)}{d \hat x}\,.
\end{eqnarray}
The derivation of the expression for the phoretic slip-velocity is
concluded by noting that $\dfrac{d \rho}{d \hat x} =
\dfrac{1}{R_+} \dfrac{d \rho}{d \theta}$ corresponds to the gradient
of the solute number density along the particle surface
[see Eqs. (\ref{slip_vel})-(\ref{lambda_def})].

\section{\normalsize Calculation of the hydrodynamic flow in the
outer region (beyond the surface film) and of the hydrodynamic force}
\label{app_B}
A general solution for the three-dimensional steady-state force-free
Stokes equations,
\begin{subequations}
\label{Stokes_no_force}
\begin{equation}
\label{St_flow}
\mu \nabla^2 \mathbf{u} = \nabla p\,,
\end{equation}
\begin{equation}
\label{div_flow}
\nabla \mathbf{u} = 0 \,,
\end{equation}
\end{subequations}
in spherical coordinates has been obtained by Lamb
\cite{Happel_book,Lamb_book}. Here we briefly present the derivation,
adapted to the present case which has an additional azimuthal
symmetry, and we also compute the hydrodynamic force in the case of a
flow subject to the boundary conditions given by Eq. (\ref{BC_flow}).

The general solution is written as $\mathbf{u} = \mathbf{u}_{hom}
+ \mathbf{u}_{p}$, where $\mathbf{u}_{hom}$ is the solution of
the homogeneous equations, i.e.,
$\mu \nabla^2  \mathbf{u}_{hom} = 0$ (so that the components
of $\mathbf{u}_{hom}$ are harmonic functions, i.e., they obey
the Laplace equation) and $\nabla \mathbf{u}_{hom} = 0$,
while $\mathbf{u}_{p}$ is a particular solution of
Eqs. (\ref{St_flow}) and (\ref{div_flow}).

\subsection{General solution in spherical coordinates in the presence
of azimuthal symmetry}

The calculation of $\mathbf{u}$ proceeds by separately determining
the components $\mathbf{u}_{hom}$ and  $\mathbf{u}_{p}$ defined
above. The construction of the component $\mathbf{u}_{p}(r,\theta)$
[a particular solution of the inhomogeneous Eq. (\ref{St_flow})]
starts from the observation that the pressure field is also a
harmonic function, i.e., $\nabla^2 p = 0$ [this follows from taking
the divergence of Eq. (\ref{St_flow})].
Therefore, it can be expanded in terms of the solid harmonics
\cite{Happel_book}, which are the eigenfunctions of the Laplace
operator in \textit{3d}:
\begin{equation}
\label{press_expansion}
p(r,\theta) = \sum_{\ell \in \mathbb{Z}}
{\tilde p}_\ell K_\ell(r,\theta)\,,
\end{equation}
where
\begin{equation}
\label{solid_harmonics}
K_\ell(r,\theta) = r^\ell Q_\ell(\cos\theta)\,,\ell \in \mathbb{Z} \,,
\end{equation}
with
\begin{equation}
\label{Legendre_P}
Q_\ell(\cos\theta) =
\begin{cases}
P_\ell(\cos\theta)\,,~\mathrm{for}~\ell \geq 0\,,\\
P_{|\ell|-1}(\cos\theta)\,,~\mathrm{for}~\ell < 0\,,
\end{cases}
\end{equation}
and $P_\ell(\cos\theta)$ is the Legendre polynomial of degree $\ell$.

In this representation, a particular solution $\mathbf{u}_{p}$ is
given by
\cite{Lamb_book,Happel_book}
\begin{equation}
\label{part_sol}
\mathbf{u}_{p} = \sum_{\ell \in \mathbb{Z}}
{\tilde p}_\ell
[A_\ell r^2 \nabla K_\ell + B_\ell \,\mathbf{r} \, K_\ell]\,
\end{equation}
where
\begin{equation}
\label{coeff_A_B}
A_\ell = \dfrac{\ell+3}{2 \mu (\ell+1)(2\ell+3)}\,,~
B_\ell = -\dfrac{2 \ell}{\ell+3} A_\ell\,.
\end{equation}
This can be checked by inserting Eqs. (\ref{part_sol}) and
(\ref{coeff_A_B}) into Eqs. (\ref{Stokes_no_force}a, b) and by
noting that $\nabla \mathbf{r} = 3$ and
$r \partial_r K_\ell = \ell K_\ell$. Note that the pole in
$A_\ell$ at $\ell = -1$ is cancelled in Eq. (\ref{part_sol}).

The construction of the solution $\mathbf{u}_{hom}$ starts from the
identity
\begin{eqnarray}
\label{vec_ident}
\nabla (\mathbf{r} \mathbf{u}_{hom}) &=&
\mathbf{r} \times (\nabla \times \mathbf{u}_{hom}) +
\mathbf{u}_{hom} \times (\nabla \times \mathbf{r}) \nonumber\\
&+& (\mathbf{u}_{hom} \cdot \nabla) \,\mathbf{r} +
(\mathbf{r} \cdot \nabla) \,\mathbf{u}_{hom}\,.
\end{eqnarray}
Expressing the vorticity $\nabla \times \mathbf{u}_{hom}$ as
$\nabla \Upsilon$ [this is possible because
$\nabla \times (\nabla \times \mathbf{u}_{hom}) =
\nabla (\nabla \mathbf{u}_{hom}) - \nabla^2 \mathbf{u}_{hom} = 0$]
and denoting the product $\mathbf{r} \cdot \mathbf{u}_{hom}$ by
$\Phi$, and noting that: (i) $ \nabla \times \mathbf{r} = 0$;
(ii) $(\mathbf{u}_{hom} \cdot \nabla) \mathbf{r} = \mathbf{u}_{hom}$;
(iii) $(\mathbf{r} \cdot \nabla) \mathbf{u}_{hom} =
r \partial_r \mathbf{u}_{hom}$; (iv)
$\mathbf{r} \times (\nabla \Upsilon)
= - \nabla \times (\mathbf{r} \Upsilon)$, Eq. (\ref{vec_ident})
can be re-written as
\begin{equation}
\label{hom_sol_form}
\mathbf{u}_{hom} + r \partial_r \mathbf{u}_{hom} =
\nabla \Phi + \nabla \times(\mathbf{r} \Upsilon)\,.
\end{equation}
Due to $\nabla \times \mathbf{u}_{hom} = \nabla \Upsilon$ one
has $ \nabla^2 \Upsilon = 0$ so that $\Upsilon$ is harmonic, and
due to $\nabla^2 \Phi = 2 \nabla \mathbf{u}_{hom} +
\mathbf{r} (\nabla^2 \mathbf{u}_{hom}) = 0$ also $\Phi$ is a
harmonic function. By inserting the series expansions in solid
harmonics [compare Eq. (\ref{press_expansion})] of
$\mathbf{u}_{hom}$, $\Upsilon$, and $\Phi$ into
Eq. (\ref{hom_sol_form}) and by using
$r \partial_r K_\ell = \ell K_\ell$, one finds that the
solution $\mathbf{u}_{hom}$ is given by
\begin{equation}
\label{hom_sol}
\mathbf{u}_{hom} = \sum_{\ell \in \mathbb Z}
[{\tilde \Upsilon}_\ell \nabla \times (\mathbf{r} K_\ell) +
{\tilde \Phi}_\ell \nabla K_\ell]\,,
\end{equation}
where ${\tilde \Upsilon}_\ell$ and ${\tilde \Phi}_\ell$ are the
corresponding expansion coefficients of $\Upsilon$ and $\Phi$ as
in Eq. (\ref{press_expansion}).

The sum of Eqs. (\ref{part_sol}) and (\ref{hom_sol}) provides the
general solution $\mathbf{u} = \mathbf{u}_{p} + \mathbf{u}_{hom}$
expressed as a series expansion in spherical harmonics
\cite{Happel_book,Lamb_book} with the coefficients
$\{\tilde{p}_\ell, \tilde{\Upsilon}_\ell, \tilde{\Phi}_\ell \}$
determined by boundary conditions \textit{on spherical surfaces}.
Note that for a problem without azimuthal symmetry the derivation of
the solution proceeds analogously and the only change is that the
Legendre polynomials $P_\ell(\cos \theta)$ ($\ell \geq 0$) are
replaced everywhere by a linear combination (with respect to the
index $- \ell \leq m \leq \ell$) of the spherical harmonics
$Y_{\ell m}(\theta,\phi)$.

\subsection{Boundary conditions for problems with spherical
symmetry}

The general solution derived in the previous subsection allows one
to determine the hydrodynamic flow obeying
Eqs. (\ref{Stokes_no_force}a, b) in a spherical domain with
prescribed velocity on the boundaries of the domain (or at
infinity). This is carried out in the usual manner by expanding the
prescribed surface fields in terms of Legendre polynomials
and then equating them with the series representation of the general
solution evaluated at the boundary to determine the expansion
coefficients $\{\tilde{p}_\ell, \tilde{\Upsilon}_\ell,
\tilde{\Phi}_\ell \}$.
However, the prescribed boundary conditions for velocity fields can
be used to obtain an equivalent set of boundary conditions, which
exploits the simplicity of the derivatives of the solid harmonics
with respect to the radial coordinate and thus significantly
simplifies the algebra \cite{Happel_book}; here we follow this
approach, and discuss the most general case, i.e., without
azimuthal symmetry.

Let $\mathbf{U}$ denote the prescribed velocity field on a
spherical boundary $|\mathbf{r}| = c$,
i.e., $\mathbf{u} (r = c,\theta,\phi) = \mathbf{U}(\theta,\phi)$;
in our case $\mathbf{U} = \mathbf{V} + \mathbf{v}_s$ for $c = R_+$
and $\mathbf{U} = 0$ for $c = R_1$.
\newline
\textbf{(i)} By multiplying $\mathbf{u}(r = c, \theta,\phi)
= \mathbf{U}(\theta,\phi)$ with $\mathbf{e}_r$, one obtains the
radial component $u_r$ of the flow field
\begin{equation}
\label{v_radial_BC}
u_r(r = c,\theta,\phi) =
\mathbf{e}_r \cdot \mathbf{U}(\theta,\phi)=:
U_r(\theta,\phi)\,,
\end{equation}
where $U_r$ denotes the radial component of the velocity
$\mathbf{U}$ prescribed at the spherical boundary.
\newline
\textbf{(ii)} The radial component of the vorticity is given by
\begin{equation}
\mathbf{e}_r \cdot (\nabla \times \mathbf{u})
= \dfrac{1}{r \sin \theta}\,
[\partial_\theta (\sin\theta \, u_\phi) - \partial_\phi u_\theta]\,,
\label{vort_radial}
\end{equation}
where $u_\theta$ and $u_\phi$ are the polar and azimuthal components
of $\mathbf{u}$, respectively. On the spherical boundary
$|\mathbf{r}| = c$, within the right hand side of
Eq. (\ref{vort_radial}) $u_\phi$ and $u_\theta$ can be
replaced by $U_\phi$ and $U_\theta$, respectively. Thus the
radial component of the vorticity obeys
\begin{eqnarray}
\label{vort_radial_BC}
[\mathbf{r} \cdot (\nabla \times \mathbf{u})]|_{r = c} &=&
\mathbf{r} \cdot (\nabla \times \mathbf{U}) \nonumber\\
&=&
\dfrac{1}{\sin \theta}\,
[\partial_\theta (\sin\theta \, U_\phi) - \partial_\phi U_\theta]
\,.\hspace*{.3in}
\end{eqnarray}
\newline
\textbf{(iii)} Since $\nabla \mathbf{u} = 0$, one has
\begin{eqnarray}
0 = r \nabla \mathbf{u} &=& r \,\partial_r u_r + 2 u_r \nonumber\\
&+&
\dfrac{1}{\sin \theta}
[\partial_\theta (\sin\theta \, u_\theta) + \partial_\phi u_\phi]\,.
\label{div_modify}
\end{eqnarray}
Since $\partial_r U_r(\theta,\phi) = 0$,
$u_\theta(r = c,\theta,\phi) = U_\theta(\theta,\phi)$, and
$u_\phi(r = c,\theta,\phi) = U_\phi(\theta,\phi)$,
Eq. (\ref{div_modify}) renders on $|\mathbf{r}| = c$ the boundary
condition
\begin{equation}
\label{div_BC}
[r \,\partial_r u_r]|_{r = c} = - r (\nabla \mathbf{U})\,.
\end{equation}

Determining $u_r$, $\mathbf{r} \cdot (\nabla \times \mathbf{u})$,
and $r \partial_r u_r$ from the solution $\mathbf{u} =
\mathbf{u}_{hom} + \mathbf{u}_{p}$ [Eqs. (\ref{part_sol})
and (\ref{hom_sol})], evaluating the expressions at $r = c$, and
equating them to the rhs of Eqs. (\ref{v_radial_BC}),
(\ref{vort_radial_BC}), and (\ref{div_BC}), respectively, after
expanding the latter in terms of spherical harmonics, yields the
equations of condition for the coefficients $\{\tilde{p}_\ell,
\tilde{\Upsilon}_\ell, \tilde{\Phi}_\ell \}$. Applying this procedure
for all boundaries, combining all the resultant conditions, and
solving for the coefficients $\{\tilde{p}_\ell, \tilde{\Upsilon}_\ell,
\tilde{\Phi}_\ell \}$  yields the velocity $\mathbf{u}$ in terms
of $\mathbf{U}(\theta,\phi) = \mathbf{u} (r = c,\theta,\phi)$.

\subsection{The outer hydrodynamic flow}

We now apply the general results discussed in the previous
subsections to the particular confined system shown in
Fig. \ref{fig1}, which exhibits azimuthal symmetry and obeys the
boundary conditions in Eq. (\ref{BC_flow}) on the spherical surfaces
$|\mathbf{r}| = R_+ \simeq R$ and $R_1 = \eta R$, respectively.
Using the expansion of $\mathbf{u} = \mathbf{u}_p +
\mathbf{u}_{hom}$ in terms of solid harmonics [Eqs. (\ref{part_sol})
and (\ref{hom_sol})] and exploiting the azimuthal symmetry (i.e., the
solution does not depend on $\phi$), splitting this series into two,
corresponding to $\ell \geq 0$ and $\ell < 0$, respectively, changing
in the latter the index of summation according to
$\ell \mapsto -(n+1)\,,~ n \geq 0$, and then renaming $n$ by $\ell$,
one obtains \cite{Happel_book}
\begin{subequations}
\label{lhs_expansions}
\begin{eqnarray}
\label{rad_vel_Leg_pol}
u_r &=& \sum_{\ell \geq 0}
\left[
\dfrac{\ell}{2 \mu (2\ell+3)} r^{\ell+1} {\tilde p}_\ell +
\dfrac{\ell+1}{2 \mu (2\ell-1) r^{\ell}} \,{\tilde p}_{-(\ell+1)}
\right.
\nonumber\\
&+&
\left.
\ell r^{\ell-1} {\tilde \Phi}_\ell -
\dfrac{\ell+1}{r^{\ell+2}} \,{\tilde \Phi}_{-(\ell+1)}
\right]
P_\ell(\cos\theta)\,,
\end{eqnarray}
\begin{equation}
\label{rad_vort_Leg_pol}
r\,(\nabla \times \mathbf{u})_r =
\sum_{\ell \geq 0} \ell (\ell+1)
\left[r^\ell {\tilde \Upsilon}_\ell +
\dfrac{{\tilde \Upsilon}_{-(\ell+1)}}{r^{\ell+1}}\right]
P_\ell(\cos\theta)\,,
\end{equation}
\begin{eqnarray}
\label{rad_div_Leg_pol}
&&r\,\partial_r u_r =
\sum_{\ell \geq 0}
\left[
\dfrac{\ell (\ell+1)}{2 \mu (2\ell+3)} r^{\ell+1} {\tilde p}_\ell -
\dfrac{\ell (\ell+1)}{2 \mu (2\ell-1) r^{\ell}} \,{\tilde p}_{-(\ell+1)}
\right.
\nonumber\\
&&+
\left.
\ell (\ell-1) r^{\ell-1} {\tilde \Phi}_\ell +
\dfrac{(\ell+1)(\ell+2)}{r^{\ell+2}} \,{\tilde \Phi}_{-(\ell+1)}
\right]
P_\ell(\cos\theta)\,.\nonumber\\
\end{eqnarray}
\end{subequations}
Equations (\ref{rad_vel_Leg_pol}), (\ref{rad_vort_Leg_pol}), and
(\ref{rad_div_Leg_pol}) relate to Eqs. (\ref{v_radial_BC}),
(\ref{vort_radial_BC}), and (\ref{div_BC}), respectively. The rhs of
Eqs. (\ref{rad_vel_Leg_pol}) and (\ref{rad_div_Leg_pol}) do not
depend on the coefficients $\{\tilde{\Upsilon}_\ell\}$ because the
contributions
$\tilde{\Upsilon}_\ell \nabla \times (\mathbf{r} K_\ell)$ in
Eq. (\ref{hom_sol}) have no radial component. Similarly, the rhs of
Eq. (\ref{rad_vort_Leg_pol}) does not depend on
$\{\tilde{p}_\ell\}$ and $\{\tilde{\Phi}_\ell\}$ because
$\mathbf{r} \cdot (\nabla \times \mathbf{u}_p) = 0$
[Eq. (\ref{part_sol})] and $\nabla \times (\nabla K_\ell) = 0$
[Eq. (\ref{hom_sol})], respectively.

By noting that $\nabla \mathbf{V} = 0$, $\nabla \times \mathbf{V}
= 0$ [because the particle velocity $\mathbf{V} = V \mathbf{e}_z$
depends only on $t$ and $\eta$, see Eq. (\ref{velocity})], and
$\nabla \times \mathbf{v}_s = 0$ [because the slip velocity is given
by a gradient, see Eq. (\ref{slip_vel})], one obtains that on the
spherical surface $|\mathbf{r}| = R_+ \simeq R$ one has [see
Eq. (\ref{v_radial_BC})] $U_r = (\mathbf{V}+\mathbf{v}_s)_r
= V \cos\theta$ [because $\mathbf{V} = V \mathbf{e}_z$ and
$\mathbf{e}_r \mathbf{v}_s = 0$, see Eq. (\ref{slip_vel_hatx})],
$\mathbf{r} \cdot [\nabla \times \mathbf{U}] =
\mathbf{r} \cdot [\nabla \times (\mathbf{V}+\mathbf{v}_s)] = 0$
[see Eq. (\ref{vort_radial_BC})], and [see Eq. (\ref{div_BC})]
\begin{equation}
\label{div_vs}
- r \nabla \mathbf{U} = - r \nabla (\mathbf{V}+\mathbf{v}_s)
= - r \nabla \mathbf{v}_s = - k(\theta,R)\,,
\end{equation}
where [due to $\mathbf{v}_s = - b \nabla_s \rho(R,\theta;t,\eta)$]
\begin{eqnarray}
\label{def_func_h}
k(\theta,R) &=& - \dfrac{b}{R \sin\theta} \,
\partial_\theta \,(\sin\theta \,\partial_\theta \rho)\nonumber\\
&=& \sum_{\ell \geq 0} {\tilde k}_\ell(R) P_\ell(\cos\theta)\,,
\end{eqnarray}
$\rho$ denotes $\rho(R,\theta;t,\eta)$, and
\begin{eqnarray}
\label{coef_h}
{\tilde k}_\ell(R)&=& \dfrac{2 \ell + 1}{2}
\int\limits_{0}^{\pi} d\theta \sin\theta\,
k(\theta,R) \,P_\ell(\cos\theta)
\nonumber\\
&=& \dfrac{2 \ell + 1}{2} \,\dfrac{b}{R}
\int\limits_{0}^{\pi} d\theta \sin\theta \,
\dfrac{\partial\rho}{\partial\theta}
\dfrac{dP_\ell(\cos\theta)}{d\theta}\,.\hspace*{.2in}
\end{eqnarray}
(The last equality follows upon integrating by parts.) By
evaluating the right hand sides of Eqs. (\ref{rad_vel_Leg_pol}),
(\ref{rad_vort_Leg_pol}), and (\ref{rad_div_Leg_pol}) at $R$ and
$R_1$ and equating them on the left hand sides with
$V \cos(\theta), 0, - k(\theta,R)$ and with 0, 0, 0, respectively,
the coefficients $\{\tilde{p}_\ell, \tilde{\Upsilon}_\ell,
\tilde{\Phi}_\ell \}$ are obtained for any $\ell \geq 0$ as the
solution of the system of equations given by
\begin{subequations}
\label{system_coef}
\begin{eqnarray}
\label{rad_vel_R}
&&\dfrac{\ell \,R^{\ell+1}}{2 \mu (2\ell+3)}  {\tilde p}_\ell +
\dfrac{\ell+1}{2 \mu (2\ell-1) R^{\ell}} \,{\tilde p}_{-(\ell+1)}
\nonumber\\
&&+
\ell R^{\ell-1} {\tilde \Phi}_\ell -
\dfrac{\ell+1}{R^{\ell+2}} \,{\tilde \Phi}_{-(\ell+1)}
= V \delta_{\ell,1}\,,
\end{eqnarray}
\begin{equation}
\label{rad_curl_R}
R^\ell {\tilde \Upsilon}_\ell +
\dfrac{{\tilde \Upsilon}_{-(\ell+1)}}{R^{\ell+1}} = 0\,,
\end{equation}
\begin{eqnarray}
\label{rad_div_R}
&& \dfrac{\ell (\ell+1)\,R^{\ell+1}}{2 \mu (2\ell+3)} {\tilde p}_\ell -
\dfrac{\ell (\ell+1)}{2 \mu (2\ell-1) R^{\ell}} \,{\tilde p}_{-(\ell+1)}
\nonumber\\
&+&
\ell (\ell-1) R^{\ell-1} {\tilde \Phi}_\ell +
\dfrac{(\ell+1)(\ell+2)}{R^{\ell+2}} \,{\tilde \Phi}_{-(\ell+1)}
= - {\tilde k}_\ell(R)\,,\nonumber\\
\end{eqnarray}
\begin{eqnarray}
\label{rad_vel_R1}
&&\dfrac{\ell \,R_1^{\ell+1}}{2 \mu (2\ell+3)}  {\tilde p}_\ell +
\dfrac{\ell+1}{2 \mu (2\ell-1) R_1^{\ell}} \,{\tilde p}_{-(\ell+1)}
\nonumber\\
&&+
\ell R_1^{\ell-1} {\tilde \Phi}_\ell -
\dfrac{\ell+1}{R_1^{\ell+2}} \,{\tilde \Phi}_{-(\ell+1)}
= 0\,,
\end{eqnarray}
\begin{equation}
\label{rad_curl_R1}
R_1^\ell {\tilde \Upsilon}_\ell +
\dfrac{{\tilde \Upsilon}_{-(\ell+1)}}{R_1^{\ell+1}} = 0\,,
\end{equation}
\begin{eqnarray}
\label{rad_div_R1}
&&\dfrac{\ell (\ell+1)\,R_1^{\ell+1}}{2 \mu (2\ell+3)} {\tilde p}_\ell -
\dfrac{\ell (\ell+1)}{2 \mu (2\ell-1) R_1^{\ell}} \,{\tilde p}_{-(\ell+1)}
\nonumber\\
&&+
\ell (\ell-1) R_1^{\ell-1} {\tilde \Phi}_\ell +
\dfrac{(\ell+1)(\ell+2)}{R_1^{\ell+2}} \,{\tilde \Phi}_{-(\ell+1)} = 0
\,.\nonumber\\
\end{eqnarray}
\end{subequations}
Note that Eqs. (\ref{rad_curl_R}) and (\ref{rad_curl_R1}) imply
that ${\tilde \Upsilon}_\ell = {\tilde \Upsilon}_{-\ell} = 0$ for
all $\ell$. Therefore, for any given $\ell$ one is left with a
linear system of four equations for four unknowns: ${\tilde p}_\ell,
{\tilde \Phi}_\ell, {\tilde p}_{-(\ell+1)},
{\tilde \Phi}_{-(\ell+1)}$, which (in the generic case) admits a
unique solution.

Taking $\ell = 1$ in Eq. (\ref{system_coef}), one obtains a
system of four linear equations for the unknowns ${\tilde p}_{1},
{\tilde \Phi}_{1}, {\tilde p}_{-2}, {\tilde \Phi}_{-2}$, which
depend parametrically on $V, R, R_1$, and $\mu$ [see Eq.
(\ref{coeff_A_B})]. The absence of a body force acting on the
particle implies ${\tilde p}_{-2}$ [see Eqs. (\ref{hydro_force}) -
(\ref{force_p2})]. This leads to:
\begin{equation}
\label{V_k1}
V = \dfrac{{\tilde k}_1(R)}{3}
\left[ 1- \dfrac{5}{2} \dfrac{\eta^2-1}{\eta^5-1}\right]\,.
\end{equation}
By noting that
\begin{eqnarray}
\label{tilde_k1}
{\tilde k}_1(R)
&=& \dfrac{3b}{2R} \int\limits_{0}^{\pi} d\theta \sin\theta \,
\dfrac{\partial\rho}{\partial\theta} \dfrac{dP_1(\cos\theta)}{d\theta}
\\
&=& - \dfrac{3b}{2R} \int\limits_{0}^{\pi} d\theta \sin^2\theta \,
\dfrac{\partial\rho}{\partial\theta}
\nonumber\\
&=& 3 \dfrac{b}{R}
\int\limits_{0}^{\pi} d\theta \sin\theta \cos\theta \rho(R,\theta)\,,
\nonumber
\end{eqnarray}
one obtains the expression Eq. (\ref{velocity}) in the main text.
Because $\partial_\theta \rho$ is positive for the system depicted
in Fig. \ref{fig1} the second equality in Eq. (\ref{tilde_k1})
implies that the sign of ${\tilde k}_1(R)$, and therefore that of
$V$ [see Eq. (\ref{V_k1})], is opposite to the one of the effective
mobility $b$.

Before concluding this appendix, we note that the general solution
for the hydrodynamic flow [Eqs. (\ref{part_sol}), (\ref{hom_sol}),
and (\ref{system_coef})] allows one to gain insight into the
qualitative differences between the phoretic motion mechanism and
the one of an ``osmotic pressure'' propeller \cite{Brady_2008}.
\begin{figure}[!htb]
\includegraphics[width=.95\linewidth]{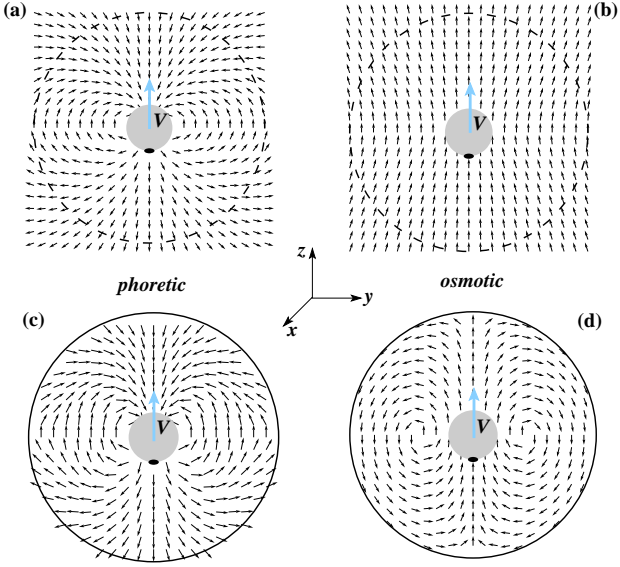}
\caption{
\label{fig4}
Normalized velocity fields $\mathbf{u}/|\mathbf{u}|$ in the
laboratory frame corresponding to the diffusiophoretic motion due to
steric repulsion only ($b < 0$) between the solute and the particle
for the case of (a) an unbounded system and (c) a confined system
($\eta = 5$). The panels (b) and (d) show the corresponding results
in the case that the motion occurs as a result of a drag provided by
the ``osmotic pressure'' integrated along the surface of the particle.
The dashed circles in (a) and (b) indicate the position at which the
confining wall is placed in (c) and (d). In all cases (a) - (d) the
particle moves upwards, i.e., its velocity $\mathbf{V}$ is in the
positive $z$-direction (as shown by the vector on the particle).
Here we consider only time scales on which the movement of the particle
does not change the flow fields.
}
\end{figure}
In Figs. \ref{fig4}(a) and (c) we show the normalized velocity
fields $\mathbf{u}/|\mathbf{u}|$ corresponding to diffusiophoretic
motion due to steric repulsion only. In this case $b < 0$
[see Eq. (\ref{lambda_def})] so that the particle velocity
$\mathbf{V} = V \mathbf{e}_z$ is oriented along the positive
$z$-direction, i.e., away from the source. Figure \ref{fig4}(a)
corresponds to an unbounded system whereas Fig. \ref{fig4}(c)
depicts the corresponding confined system with $\eta = 5$; both
cases refer to the same (chosen) constant surface gradient
$\partial_\theta \rho(R,\theta;\eta) > 0$ of the solute
number density. (Being a constant, this gradient scales out and
is absorbed in the velocity scale. Therefore the dimensionless
ratio $\mathbf{u}/|\mathbf{u}|$ is independent of it.) Since $b$
enters the problem only via the multiplication of
$\partial_\theta \rho(R,\theta;t,\eta)$ [see Eq. (\ref{tilde_k1})],
also the absolute value of $b$ drops out from the normalized flow
field. The viscosity $\mu$ drops out, too. For the unbounded case,
the solution [see also Eq. (35) in Ref. \onlinecite{Anderson_1989}] is
determined from Eqs. (\ref{rad_vel_R}), (\ref{rad_curl_R}), and
(\ref{rad_div_R}) and the requirement that the flow field vanishes
at infinity. This implies that in Eq. (\ref{lhs_expansions}) all
coefficients multiplying terms $\sim r^\ell$ with $\ell \geq 0$ are zero.
In the confined case, the flow field is approximated by keeping
terms up to $\ell = 50$ [Eqs. (\ref{system_coef})] in the series
expansion of Eqs. (\ref{part_sol}) and (\ref{hom_sol}); this
provides a reasonably good approximation, except near the boundary
at $R_1$ and for polar angles close to zero or $\pi$ where
apparently there are deviations from the no-slip boundary
condition. However, one should keep in mind that in this
normalized representation of the flow $\mathbf{u}/|\mathbf{u}|$ small
numerical deviations from $\mathbf{u} = 0$ at $R_1$ are sharply
overemphasized. For comparison, in Figs. \ref{fig4}(b) and (d) we
present the corresponding results for the case in which the
motion of the particle occurs as a result of a drag provided
by an ``osmotic pressure'' integrated over the surface of the
particle. In this case the particle velocity is oriented along
the positive $z$-direction, too, because of the assumption that
the pressure is proportional to the solute number density and
thus is larger at the lower hemisphere. In this case Eq.
(\ref{system_coef}) is solved under the assumption of a no-slip
boundary condition at $R$, i.e., all the coefficients
$\tilde{k}_\ell$ are set to zero; this allows one to obtain a
solution in closed form in both the unbounded and the confined
cases. Note that in this case $p_{-2}$ is non-zero. It is
fixed by the given drag [integrated ``osmotic pressure'',
Eq. (\ref{force_p2})], and thus Eq. (\ref{system_coef}), evaluated
at $\ell =  1$, leads to a relation between the velocity $V$ and
the drag force. (In the unbounded case, this relation is the
well-known Stokes formula for the viscous friction on a sphere.)
In Eq. (\ref{rad_vel_R}) $V$ sets the velocity scale, which drops
out (as well as the drag force, because $V$ is proportional to it)
of the dimensionless ratio $\mathbf{u}/|\mathbf{u}|$.

Qualitative differences between these two mechanisms are particularly
clearly visible in the case of the unbounded systems. For example,
the flow along the $z$-axis is in the negative direction (i.e.,
opposite to the motion of the particle) in the phoretic case, while
it is in the positive direction for the ``osmotic'' propeller; the
flow lines for the phoretic motion are closing back and then
``slide'' along the surface of the particle. Since closed-form
analytical expressions are available in these unbounded cases
\cite{Anderson_1989,Happel_book}, the qualitative differences noticed
above can be correlated with the different characteristics of the
flows. In the case of phoretic motion the flow field decays as
$r^{-3}$ at large distances $r$ from the particle, which is a
``multipole'' far-field corresponding to the action of a
force-quadrupole disturbance, in contrast to the $r^{-1}$ decay in
the osmotic case, which is a ``monopole'' far-field corresponding to
the action of a point-force (``Stokeslet'') disturbance. The faster
decay of the velocity field in the first case means that a boundary
at $R_1$ will perturb the phoretic flow much less. This explains why
for the phoretic motion the viscous friction increases with the
confinement weaker than what would be expected from naively using a
Stokes-friction concept.

The flow fields remain qualitatively different even in the presence
of the confinement (in particular with respect to the opposite
direction of flow along the $z$-axis), although the differences are
not as pronounced as in the unbounded case. A further qualitative
difference between the two confined flows is that in (d) a vortex
structure is formed fully detached from the particle whereas in (c)
the vortex involves flow along the surface of the particle. Due to
the presence of the boundary at $R_1$, all the terms $\sim r^\ell$
in the expansions of Eq. (\ref{lhs_expansions}) are present in the
solution and an analysis of the flow in terms of fundamental
solutions, as for the unbounded case, is no longer possible. Note
that since $\nabla \times \mathbf{u}_{hom} = \nabla \Upsilon$ and
$\Upsilon = 0$ [see the text following Eqs. (\ref{vec_ident}) and
(\ref{rad_div_R1})] one has $\nabla \times \mathbf{u} =
\nabla \times \mathbf{u}_p$ so that with Eqs. (\ref{part_sol}) and
(\ref{coeff_A_B}) one obtains
\begin{equation}
\nabla \times \mathbf{u} = \dfrac{1}{\mu} \,\mathbf{e}_\phi
\sum_{\ell \in \mathbb{Z}} \dfrac{{\tilde p}_\ell}{\ell + 1}
\dfrac{\partial K_\ell(r,\theta)}{\partial \theta}\,,
\end{equation}
which is in general nonzero [e.g., for ``osmotic" flow
${\tilde p}_{-2} \neq 0$ (Eq. (\ref{force_p2}))]. Thus in agreement
with Fig. \ref{fig4} the flow field can contain vortices.

\section{Calculation of the number density of product molecules}
\label{app_C}

Equation (\ref{diff_eq}), subject of the IC and BC conditions given
in Eq. (\ref{BC_IC}) is solved by using the Laplace transform and
the inversion theorem \cite{Carslaw_book}:
\begin{equation}
\bar f(\zeta) \equiv {\cal L} [f]
=  \int\limits_0^\infty dt \,e^{-\zeta t} f(t)
\label{LT}
\end{equation}
and
\begin{equation}
f(t)\equiv {\cal L}^{-1}[\bar f(\zeta)]:=
\frac{1}{2\pi i}
\int\limits_{\gamma-i \infty}^{\gamma + i \infty}
d\zeta \,e^{\zeta t}\bar f(\zeta),
\label{ILT}
\end{equation}
respectively, where $\bar f(\zeta)$ (the Laplace transformed
quantities are indicated by an overbar) is assumed to be
well-defined for $\zeta \in \mathbb{R}^+$ and $\gamma \in \mathbb{R}$
is sufficiently large such that all singularities of $\bar f(\zeta)$
lie to the left of the integration path. By taking the Laplace
transform of Eqs. (\ref{diff_eq}) and (\ref{BC_IC}) and by using the
IC condition of zero number density of the product molecules, one
finds that the Laplace transformed density is given by
\begin{equation}
\bar \rho(\mathbf{r}, \zeta)
= - \bar a(\zeta) \bar G(\mathbf{r},q)\,,
\label{LT_dens}
\end{equation}
where $\bar a(\zeta) = \bar B(\zeta)/D$ and $\bar G(\mathbf{r},q)$
is the Green's function for the Helmholtz operator $\nabla^2 - q^2$,
$q = \sqrt{\zeta/D} > 0$, satisfying the BCs of vanishing normal
derivative at $|\mathbf{r}|= R, R_1$.

Decomposing $\bar G(\mathbf{r},q)$ as $\bar G(\mathbf{r},q) =
G_s(\mathbf{r},q;{\mathbf{r}}_s)+ g(\mathbf{r},q)$, where the
singular part $G_s$ is the free space Green's function for the
Helmholtz operator (which is known in any spatial dimension $d$, see,
e.g., Ref. \onlinecite{Hassani_book}), the initial problem is reduced to
that of finding the solution $g$ of the homogeneous Helmholtz
equation subject to the boundary conditions
\begin{equation}
\label{BC_for_g}
\left.\left(
\frac{\partial g(\mathbf{r},q)}{\partial r}
\right)\right|_{|\mathbf{r}|= R, R_1}
= -\left.\left(
\frac{\partial G_s(\mathbf{r},q;{\mathbf{r}}_s)}{\partial r}
\right)\right|_{|\mathbf{r}|= R, R_1}\,.
\end{equation}

In \textit{3d} the singular part of the Green's function is
given by \cite{Hassani_book}
\begin{equation}
\label{Gs_3d}
G_s(\mathbf{r},q;{\mathbf{r}}_s) = -\frac{q}{4\pi}
\frac{e^{-q |\mathbf{r}-{\mathbf{r}}_s|}}
{q |\mathbf{r}-{\mathbf{r}}_s|},
\end{equation}
while the regular part (i.e., the general solution of the
homogeneous Helmholtz equation) can be written as
\begin{equation}
\label{g_3d}
g(r,\theta,q)
= \sum\limits_{\ell \geq 0}
[\alpha_\ell \, i_{\ell+1/2}(q r)
+ \beta_\ell \, k_{\ell+1/2}(q r)]
P_\ell(\cos\theta),
\end{equation}
where $i_{\ell+1/2}(z) = \sqrt{\pi/(2 z)} I_{\ell+1/2}(z)$ and
$k_{\ell+1/2}(z) = \sqrt{\pi/(2 z)} K_{\ell+1/2}(z)$ are the
modified spherical Bessel functions of the first and third kind,
respectively \cite{Abramowitz_book}, $P_\ell$ is the Legendre
polynomial of degree $\ell$, while the coefficients
$\alpha_\ell$ and $\beta_\ell$ will be fixed to fulfill the
boundary conditions.

Using one of the addition theorems for the Bessel
functions \cite{Carslaw_book} and noting that the angle between
$\mathbf{r}$ and ${\mathbf{r}}_s$ is $\pi - \theta$
(see Fig.~\ref{fig1}), $G_s$ can be re-written as
\begin{equation}
\label{Gs_3d_expand}
G_s(\mathbf{r},q;{\mathbf{r}}_s) = -
\sum\limits_{\ell \geq 0} c_\ell(q)
i_{\ell+1/2}(q r_<) k_{\ell+1/2}(q r_>) P_\ell(\cos\theta)
\end{equation}
where $c_\ell(q) = (-1)^\ell (2\ell+1) q / (2\pi^2)$,
$r_< = \min(r,r_s)$, and $r_> = \max(r,r_s)$. By combining
Eqs. (\ref{g_3d}, \ref{Gs_3d_expand}) and the boundary conditions
[Eq. (\ref{BC_for_g})], by noting that $r_s = R + \epsilon\,,~
\epsilon \searrow 0$  (i.e., the source of particles is
\textit{on} the surface of the particle), and by re-writing the
coefficients as
$\alpha_\ell \equiv c_\ell i_{\ell+1/2}(q R) \tilde \alpha_\ell$
and $\beta_\ell \equiv c_\ell i_{\ell+1/2}(q R) \tilde \beta_\ell$,
the dimensionless coefficients $\tilde \alpha_\ell(qR)$ and
$\tilde \beta_\ell(qR)$ are determined by the solution of the
following closed system of two linear equations with two unknowns
($\tilde \alpha_\ell$ and $\tilde \beta_\ell$):
\begin{subequations}
\label{alpa_beta_3d_eqs}
\begin{eqnarray}
\label{alpa_3d}
\tilde \alpha_\ell + \tilde \beta_\ell \,w_{\ell+1/2}(q R)
&=& v_{\ell+1/2}(q R),\label{BC_at_R}\\
\tilde \alpha_\ell + \tilde \beta_\ell \,w_{\ell+1/2}(\eta q R)
&=& w_{\ell+1/2}(\eta q R),\label{BC_at_R1}
\end{eqnarray}
\end{subequations}
where $v_{\ell+1/2}(z):= k_{\ell+1/2}(z)/i_{\ell+1/2}(z)$ and
$w_{\ell+1/2}(z):= [d k_{\ell+1/2}(z)/dz]/[d i_{\ell+1/2}(z)/dz]$.
The solution ${\hat \alpha}_\ell$ and ${\hat \beta}_\ell$ acquires
a dependence on $\eta$ via Eq. (\ref{BC_at_R1}): ${\hat \alpha}_\ell
= {\hat \alpha}_\ell(qR, \eta)$ and ${\hat \beta}_\ell
= {\hat \beta}_\ell(qR, \eta)$, with $q R = R \sqrt{\zeta/D}$.
This determines the Green's function $\bar G(\mathbf{r},q)$ and
thus the Laplace transformed density $\bar \rho(\mathbf{r}, \zeta)$
[Eq. (\ref{LT_dens})] from which, in principle, the density
$\bar \rho(\mathbf{r}, t)$ is obtained via the inverse Laplace
transformation. Note that although the final result is in the
form of a series, the quantities in which we are interested, in
particular the phoretic velocity $V(t)$ [Eq. \ref{velocity}],
involve integrals over the polar angle $\theta$ of the product
between this series and a specific $\ell$ Legendre polynomial
$P_\ell(\cos\theta)$ [$\ell = 1$ in the case of the velocity,
$\ell = 0$ in the case of the total number of product molecules
(see below)] and thus only one of the terms from the series will
contribute.

We make the following three remarks:\\
\textbf(i) Since for $q R > 0$ one has
${\displaystyle \lim_{\eta \to \infty}} w_{\ell+1/2}(\eta q R) = 0$,
in the limit $\eta \to \infty$ one finds $\bar \alpha_\ell = 0$ and
Eq. (\ref{alpa_3d}) reduces to the corresponding BC in
Ref.~\onlinecite{Golestanian_2005}, i.e., as expected the solution for the
unbounded case is recovered.\\
\textbf(ii) Equations (\ref{BC_at_R}) and (\ref{BC_at_R1}) will not
coincide in the limit $\eta \to 1$ because even in this limit of
extreme confinement a source singularity is present at $r_s$ and
is picked up by $G_s(q)$ [Eq. (\ref{Gs_3d_expand})].\\
\textbf(iii) The Laplace transform $\bar N(p)$ of the total number
of product particles in the system at time $t$ is
\begin{eqnarray}
\label{N_3d}
&&\bar N(p = D q^2) = 2\pi\int\limits_{R}^{R_1} dr r^2
\int\limits_{0}^{\pi} d\theta \sin\theta
P_0(\cos\theta) [-\bar a(p)] \bar G(q)\nonumber\\
&&= - 4 \pi \bar a(p) c_0(q) i_{1/2}(q R) \nonumber\\
&& \times \int\limits_{R}^{R_1} dr r^2
[\bar \alpha_0 i_{1/2}(q r) + (\bar \beta_0 - 1) k_{1/2}(q r)] \\
&&= D \bar a(p) /p \Rightarrow N(t) =
\int\limits_{0}^{t} dt' A(t')\,, \nonumber
\end{eqnarray}
i.e., as expected $N(t)$ is given by the time integral of the
production rate, providing a welcome consistency check.


\begin{thebibliography}{0}

\bibitem{Paxton_2005}
W.F. Paxton, A. Sen, and T.E. Mallouk,
Chem.--Eur. J. \textbf{11}, 6462 (2005).
%
\bibitem{Paxton_2006}
W.F. Paxton, S. Sundararajan, T.E. Mallouk, and A. Sen,
Angew. Chem., Int. Ed. \textbf{45}, 5420 (2006);
W.F. Paxton, K.C. Kistler, C.C. Olmeda, A. Sen,
S.K.St. Angelo, Y. Cao, T.E. Mallouk, P.E. Lammert,
and V.H. Crespi,
J. Am. Chem. Soc. \textbf{126}, 13424 (2004).
%
\bibitem{Whitesides_2002}
R.F. Ismagilov, A. Schwartz, N. Bowden, and G.M. Whitesides,
Angew. Chem., Int. Ed. \textbf{41}, 652 (2002).
%
\bibitem{Sen_2005}
J.M. Catchmark, S. Subramanian, and A. Sen,
Small \textbf{1}, 1 (2005).
%
\bibitem{Golestanian_2005}
R. Golestanian, T.B. Liverpool, and A. Ajdari,
Phys. Rev. Lett. \textbf{94}, 220801 (2005).
%
\bibitem{Howse_2007}
J.R. Howse, R.A.L. Jones, A.J. Ryan, T. Gough, R. Vafabakhsh,
and R. Golestanian,
Phys. Rev. Lett. \textbf{99}, 048102 (2007).
%
\bibitem{Kapral_2007}
G. R\"uckner and R. Kapral,
Phys. Rev. Lett. \textbf{98}, 150603 (2007).
%
\bibitem{Golestanian_2007}
R. Golestanian, T.B. Liverpool, and A. Ajdari,
New J. Phys. \textbf{9}, 126 (2007).
%
\bibitem{Peruani_2007}
F. Peruani and L.G. Morelli,
Phys. Rev. Lett. \textbf{99}, 010602 (2007).
%
\bibitem{Saffman_1975}
P.G. Saffman and M. Delbr{\"u}ck,
Proc. Nat. Acad. Sci. USA \textbf{72}, 3111 (1975);
P.G. Saffman,
J. Fluid Mech. \textbf{73}, 593 (1976).
%
\bibitem{Joanny_1999}
C. Barentin, C. Ybert, J.-M. di Meglio, and J.-F. Joanny,
J. Fluid Mech. \textbf{397}, 331 (1999);
C. Barentin, P. Muller, C. Ybert, J.-F. Joanny, and J.-M. di Meglio,
Eur. Phys. J. E \textbf{2}, 153 (2000).
%
\bibitem{Anderson_1989}
J.L. Anderson,
Ann. Rev. Fluid Mech. \textbf{21}, 61 (1989).
%
\bibitem{Phibbs_1951}
M.K. Phibbs and P.A. Gigu{\`e}re, 
Can. J. Chem. \textbf{29}, 173 (1951).
%
\bibitem{S-E relation} A. Einstein,
``On the Movement of Small Particles Suspended in a Stationary
Liquid Demanded by the Molecular-Kinetic Theory of Heat'' in
\textit{Investigations on the theory of the Brownian motion},
Ed. R. F\"urth, transl. by A. D. Cowper (Dover, New York,
1956).
%
\bibitem{Prost} F. Juelicher and J. Prost, arXiv:0812.2924v1 (2008).
%
\bibitem{deGroot_book}
S.R. de Groot and P. Mazur,
\textit{Non-Equilibrium Thermodynamics}
(North-Holland, Amsterdam, 1962),
Ch. V.2, p. 44.
%
\bibitem{Ajdari_2006}
A. Ajdari and L. Bocquet,
Phys. Rev. Lett. \textbf{96}, 186102 (2006).
%
\bibitem{Happel_book}
J. Happel and H. Brenner,
\textit{Low Reynolds number hydrodynamics} (Noordhoff
International, Leyden, 1973), Ch. 3-2, pp. 62-67.
%
\bibitem{Lamb_book}
H. Lamb,
\textit{Hydrodynamics} (Dover, New York, 1945), p. 594.
%
\bibitem{Zydney_1995}
A.L. Zydney,
J. Colloid Interface Sci. \textbf{169}, 476 (1995).
%
\bibitem{Morrison_1970}
F.A. Morrison Jr.,
J. Colloid Interface Sci. \textbf{34}, 210 (1970).
%
\bibitem{Brady_2008}
U.M. C{\'o}rdova-Figueroa and J.F. Brady,
Phys. Rev. Lett. \textbf{100}, 158303 (2008).
%
\bibitem{Saidulu_2008}
N. Bala Saidulu and K. L. Sebastian,
J. Chem. Phys. \textbf{128}, 074708 (2008).
%
\bibitem{Chwang_1975}
A.T. Chwang and T.Y.-T. Wu,
J. Fluid Mech. \textbf{67}, 787 (1975).
%
\bibitem{Abramowitz_book}
M. Abramowitz and I.A. Stegun,
\textit{Handbook of Mathematical Functions with Formulas,
Graphs, and Mathematical Tables}
(Dover, New York, 1965)
pp. 374, 443.
%
\bibitem{Carslaw_book}
H.S. Carslaw and J.C. Jaeger,
\textit{Conduction of Heat in Solids}
(University Press, Oxford, 1959), pp. 377, 381.
%
\bibitem{Golestanian_2008}
P. Tierno, R. Golestanian, I. Pagonabarraga, and F. Sagu\'es,
Phys. Rev. Lett. \textbf{101}, 218304 (2008).
%
\bibitem{Hassani_book}
S. Hassani,
\textit{Mathematical Physics:
A Modern Introduction to Its Foundations}
(Springer, New York, 1998) p. 631.

\end{thebibliography}
\end{document}